\documentclass[12pt]{article}
\usepackage[left=2cm, right=2cm, bottom=2cm ,top = 2cm, footskip=1cm]{geometry}
\usepackage[font=small,labelfont=bf]{caption}

\usepackage{setspace}
\usepackage{lineno}

\let\counterwithin\relax
\usepackage{chngcntr}
\usepackage{booktabs}
\usepackage{hyperref}
\usepackage[square,numbers]{natbib}
\setcitestyle{open={(},close={)}}

\usepackage{graphicx}
\usepackage{amsmath}

\usepackage[section]{placeins}

\title{Diminished circadian and ultradian rhythms of human brain activity in pathological tissue \textit{in vivo}}
\author{
Christopher Thornton$^{1,o}$, Mariella Panagiotopoulou$^{1,o}$, \\
Fahmida A Chowdhury$^{3}$, Beate Diehl$^{3}$, John S Duncan$^{3}$,\\
Sarah J Gascoigne$^{1}$, Guillermo Besne$^{1}$, Andrew W McEvoy$^{3}$, Anna Miserocchi$^{3}$,\\
Billy C Smith$^{1}$, Jane de Tisi$^{3}$, Peter N Taylor$^{1,2,3}$,\\
Yujiang Wang$^{1,2,3,*}$}

\begin{document}

\maketitle

% author affiliations
\begin{enumerate}
\item{CNNP Lab (www.cnnp-lab.com), Interdisciplinary Computing and Complex BioSystems Group, School of Computing, Newcastle University, Newcastle upon Tyne, United Kingdom}
\item{Faculty of Medical Sciences, Newcastle University, Newcastle upon Tyne, United Kingdom}
\item{UCL Queen Square Institute of Neurology, Queen Square, London, United Kingdom}
\end{enumerate}

\begin{center}
* Yujiang.Wang@newcastle.ac.uk   \\
o Joint first authors
\end{center}

\newpage
\section*{Abstract}

Chronobiological rhythms, such as the circadian rhythm, have long been linked to neurological disorders, but it is currently unknown how pathological processes affect the expression of biological rhythms in the brain.

Here, we use the unique opportunity of long-term, continuous intracranially recorded EEG from 38 patients (totalling 6338 hours) to delineate circadian (daily) and ultradian (minute to hourly) rhythms in different brain regions. We show that functional circadian and ultradian rhythms are diminished in pathological tissue, independent of regional variations. We further demonstrate that these diminished rhythms are persistent in time, regardless of load or occurrence of pathological events.

These findings provide evidence that brain pathology is functionally associated with persistently diminished chronobiological rhythms \textit{in vivo} in humans, independent of regional variations or pathological events. Future work interacting with, and restoring, these modulatory chronobiological rhythms may allow for novel therapies.

\newpage

\section{Introduction}
Physiological processes are often modulated and structured by chronobiological rhythms. The daily, or circadian rhythm is perhaps the most pertinent and well-studied rhythm with a central pacemaker in the suprachiasmatic nucleus. Such rhythms are also differentially expressed in a range of tissues and organs, entrained by the central pacemaker. For example, multiple organs have been shown to display their own circadian rhythm in terms of gene expression and translation even in the absence of the pacemaker \citep{Hughey2016}. Even within a single organ, such as the brain, circadian regulation of gene expression is tissue-specific \cite{panda2002,mure2018}.  This local, or tissue-level, autonomous rhythmicity allows tissue-specific adaptations in phase or magnitude \citep{Yeung2018}, whilst being coordinated by a central pacemaker. Other chronobiological rhythms also exist on longer and shorter timescales, termed infradian and ultradian rhythms, respectively. However, these are far less well-studied, appear to differ in magnitude and period between individuals, organs, and tissues, and their biological mechanisms are elusive. So far, candidate drivers of some ultradian rhythms have been proposed, such as the pulsatile cortisol secretion in mammals every 1-3 hours \citep{Henley2009, Hartmann1997}, where various organs and tissues may react differently \citep{George2017, Rozhkova2021}. Taken together, there is evidence that, with or without a central pacemaker, circadian and ultradian rhythms are differentially expressed at the tissue level to temporally structure and modulate local physiological processes.

Disrupted chronobiological rhythms are often associated with dysfunction and disease. Associations between altered daily/circadian rhythms and neurodevelopmental disorders, mood disorders, epilepsy, Parkinson’s, and dementia have been reported \citep{Logan2019, Hartmann1997}. Similarly, disrupted ultradian rhythms may also be associated with neurological disorders, although their underpinning mechanisms are only beginning to be explored in animal models \citep{Monje2017, Basu2021}. Many of the reported associations are linked to behavioural disruptions on circadian and ultradian timescales (e.g. sleep, levels of physical activity). However, tissue level regulation and expression of chronobiological rhythms in brain activity is less well-studied. One ex-vivo study with 16 patients with drug-refractory temporal lobe epilepsy showed that the expression of the circadian clock gene \emph{Bmal1} is reduced in surgically resected pathological tissue compared to healthy tissue \citep{Wu2021}. Similarly, tissue-level disruption to clock gene transcription and translation are found in various animal models of epilepsy \cite{chan2021}.

Thus, emerging evidence hints at tissue-level alteration of chronobiological rhythms in pathological tissue in brain disorders. However, to date, it is completely unknown if pathological tissue shows different expression of chronobiological rhythms in brain activity and electrophysiology \textit{in vivo} in humans.

Here, we therefore investigate the rhythmicity of human brain activity \textit{in vivo}. EEG is a commonly used method to measure brain activity patterns and multiple studies have shown that rhythms on circadian and multiple ultradian timescales can also be captured as the rhythmic modulation of particular signal properties (e.g. \citep{Polich1995, Ravden1999, Polich1997, Philippu2019,Philippu2016, panagiotopoulouFluctuationsEEGBand2022}). This includes clear evidence of rhythms in the canonical EEG frequency bands following a circadian timescale \citep{Tan2003}. In contrast to e.g. gene expression data, EEG data has the clear advantage that is can be sampled continuously at a very high rate, allowing for the application of robust time series methods, such as Fourier analysis \citep{Refinetti2007}, wavelet analysis \citep{Leise2015}, and empirical mode decomposition \citep{panagiotopoulouFluctuationsEEGBand2022}. In this study, we will use intracranially recorded continuous EEG (iEEG) over multiple days to measure circadian and ultradian rhythms in humans \textit{in vivo}. Intracranial recordings have the advantage of a high signal-to-noise ratio, and provide excellent spatial resolution, as the electrodes directly sample from the target brain tissue without interference. This allows us to delineate functional chronobiological rhythms of brain activity in specific locations. 

In the following sections we will examine (1) if circadian and ultradian rhythms are altered in pathological tissue, (2) if the alterations can be attributed to pathology alone, and (3) if any alterations are dependent on seizure occurrence.

\section{Results}
\subsection{Overview of data and analysis}
We retrospectively analysed long-term iEEG recordings from 38 individuals with refractory focal epilepsy from the National Hospital for Neurology and Neurosurgery (Table \ref{tab:tab1}). These recordings were performed as part of the pre-surgical evaluation in these patients, and this retrospective study does not explicitly account for environmental factors/drivers of ultradian and circadian rhythms.  Therefore, we only use the terms ultradian and circadian here to refer to the timescale of the rhythms we observe in the iEEG data, and do not imply any environmental conditions with these terms.

From the iEEG recordings of each patient, we localised each recording contact to an anatomical region of interest (ROI) and calculated the relative band power in the delta range (1 - 4 Hz) and the other four frequency bands ( theta : 4 - 8 Hz, alpha : 8 - 13 Hz, beta : 13 - 30 Hz and gamma : 30 - 47.5 Hz, 52.5 - 57.5 Hz, 62.5 - 77.5 Hz) using a non-overlapping 30 second window over the continuous recording in all patients (recording duration between 2-21 days). Circadian (a period of 19 - 31 hours) and ultradian (1-3h, 3-6h, 6-9h, 9-12h, 12-19h periods) rhythms were then isolated by applying band pass filters to the signal of relative band power over days in each ROI.

Clinically identified regions where seizures originate (seizure onset zone, or SOZ) are often used to approximate areas that are hypothesised to cause or support seizures as part of a network. These regions are often also associated with some histopathological finding when resected. We therefore propose to use SOZ regions as a proxy for pathological tissue. To ensure robustness of our results, we also used the alternative definition of the tissue that was subsequently surgically removed, as it was deemed central to generating epileptic seizures. We obtained similar results (Suppl.~\ref{suppl:resvsspared}) with either definition, and will use the broad term of pathological tissue throughout the paper to refer to tissue that is most likely functionally (causing seizures) and/or structurally pathological.

\subsection{Circadian rhythms of brain activity are diminished in pathological tissue}
Figure~\ref{fig:fig1} (a) shows the power of the delta EEG band over seven days for an example patient, filtered to isolate only the circadian rhythm (19h-31h). In this patient the regions with the weakest circadian rhythm are within the left inferior parietal lobe, visualized in Figure~\ref{fig:fig1} (b). These comprise two out of the three pathological regions. Figure~\ref{fig:fig1} (c) compares the power of the delta circadian rhythm in regions with pathology against the remaining regions. We quantify this difference using the non-parametric Area Under the Curve (AUC) of a Receiver Operating Characteristics analysis \citep{Fawcett2006}. Here, AUC values greater than 0.5 imply diminished power in pathological tissue and the example patient yielded an AUC=0.92. Figure~\ref{fig:fig1} (d) shows the AUC values across all 38 patients - the distribution is substantially greater than 0.5 (median AUC=0.62, p=0.005) implying a diminished circadian rhythm in delta power in the pathological tissue across a majority of patients (66\% of subjects with AUC$>$0.5). When we perform the same analysis for the other canonical EEG bands (Theta, Alpha, Beta, Gamma) we find similar results (supplementary table \ref{suppl:AUCtable}).  We did not find noteworthy modulatory effects when investigating the role of age, sex, anti-seizure medication and epilepsy type. These results are shown in supplementary figures \ref{fig:agesexep} and \ref{fig:drugs}. Applying a more narrow definition of circadian rhythms (20-26 hours) did not substantially alter the distribution of AUCs (median AUC = 0.6) - shown in supplementary figure \ref{fig:circ26}. In the example subject we see lower cycle amplitude at the start and end of the recording, a feature present in a minority of cases across the cohort. Patterns of cycle amplitude across the recording are explored in supplementary section \ref{suppl:over_time}.

\subsection{Multiple ultradian rhythms are diminished in pathological brain tissue}
Figure~\ref{fig:fig2} (a) shows the delta band power, filtered to isolate the 3 - 6 hour ultradian rhythm, across regions for the same example patient. As with the circadian rhythm, there was a diminished power in the three pathological regions, quantified with an AUC of 0.96 for this patient and visualized in Figure~\ref{fig:fig2} (b). Diminished ultradian rhythms are present across the cohort, with a median AUC of 0.69 for the 3 - 6 hour rhythm, and median AUC values above 0.5 for all rhythm periods - shown in Figure~\ref{fig:fig2} (c), p$<$0.05 for all. When we perform the same analysis for the other canonical EEG bands (Theta, Alpha, Beta, Gamma) we again find similar results (supplementary table \ref{suppl:AUCtable}).

In Suppl.~\ref{suppl:rawmag} we further show that these effects in circadian and ultradian rhythms are not simply due to a diminished magnitude of the raw EEG signal in the pathological tissue.

\subsection{Pathology is independently associated with diminished chronobiological rhythms}

To investigate if our observed effects are potentially explained by the spatial location, or the broad brain lobe (or subcortical region) of the observations, we plotted the power of an example ultradian rhythm in delta grouped by lobe across patients (Figure~\ref{fig:fig3} a). While there is some variation of power of this ultradian rhythm across lobes, the power in the pathological tissue in each area appears consistently diminished. 

To show that pathology is independently associated with diminished chronobiological rhythms, we tested the association in a mixed effects regression with the lobe (those shown in figure~\ref{fig:fig3} (a)) as a covariate. This model took the power of the rhythm in each ROI as the dependent variable, the lobe of the ROI and whether the ROI has pathology (is within the SOZ) as independent variables, and grouped by patient. Figure \ref{fig:fig3} (b) shows the partially standardized fixed effect for pathology ($\beta_1$) for each ultradian/circadian rhythm in delta band power. All coefficients are below zero, indicating that ultradian/circadian rhythms are diminished in pathological tissue. Applying a likelihood ratio test to compare a model with pathology and brain area to a model with brain area only, we find that the effect of pathology explains the power of cycles significantly better for almost all EEG bands and ultradian/circadian rhythms (see Suppl \ref{suppl:AUCtable} for details).

Finally, we also checked if any particular type of pathology was associated with diminished rhythmicity, and found that most subjects showed this effect without evidence for the influence of the type of pathology (Suppl.~\ref{suppl:pathtype}).

\subsection{Diminished rhythms are persistent in time and independent of seizure occurrence}
To assess whether the rhythms of brain activity remained diminished in pathological tissue consistently over time and not only during the period around epileptic seizures (peri-ictal period), we calculated a rolling median of the AUC using a window proportional to the rhythm period multiplied by 1.5 (e.g. 36 hours for the circadian rhythm). We illustrate the circadian rhythm of delta power in our sample patient in figure \ref{fig:fig4} (a), the rolling AUC is shown in figure \ref{fig:fig4} (b) along with a histogram showing that the distributions of AUC values are similar during inter-ictal and peri-ictal periods.  Figure \ref{fig:fig4} (c) aggregates results across subjects and rhythms, showing no difference in the AUC distributions of inter-ictal and peri-ictal periods. The overall seizure load (average number of seizures per hour) was also not correlated with the AUC for the circadian rhythm of delta ($r=-0.04$) and no strong correlation was found for any other rhythm ($r<0.3$, see Suppl.~\ref{suppl:szload}). An similar analysis of only the post-ictal period found similar AUC distributions (supplementary section \ref{suppl:post_ictal} and the phase preference of seizures across the cohort is characterised in supplementary section \ref{suppl:phasepref}.

\section{Discussion}

Analysing long-term iEEG recordings from 38 subjects with refractory focal epilepsy, we identified rhythms in the power of canonical EEG bands at circadian and ultradian timescales. We found that these rhythms were diminished in pathological tissue, and that this association remained when we controlled for brain area. We also found that these rhythms remain diminished persistently across time, without dependence on seizure occurrence. Our work provides initial evidence of an association between the strength of chronobiological rhythms and the healthy functioning of brain tissue in humans \textit{in vivo}.

Circadian clock gene dysfunction has been seen in pathological tissue in multiple animal models \cite{chan2021} and human tissue \cite{Wu2021}, but a key advance in our study is to also demonstrate a functional abnormality in pathological tissue in terms of electrophysiological function in humans. From animal models we have learned that circadian gene oscillation pattern do not necessarily translate to similar changes in mRNA or protein level \cite{chan2021}. It was therefore also unclear if disruptions in these patterns translate to electrophysiological dysfunction, particularly in humans. Although numerous studies have reported circadian and ultradian rhythms in a range of signal properties in human EEG, these have been reported both in healthy participants \cite{aeschbach1999two,tan2003circadian,croce2018circadian}, as well as in e.g. epilepsy patients \cite{mitsisFunctionalBrainNetworks2020, panagiotopoulouFluctuationsEEGBand2022}. It was therefore unknown if (i) there was a dysfunction in circadian and ultradian rhythms in electrophysiology associated with pathology in humans and (ii) if so, how the dysfunction would be expressed. Our work therefore serves as crucial evidence that (i) dysfunction is seen in pathology in circadian and ultradian rhythms in electrophysiological function, and (ii) dysfunction is expressed as a persistently diminished rhythm.

Electrophysiological function is, of course, not limited to the signal features of relative band power that we analyzed. Indeed, there is evidence that other EEG signal features may display different rhythms \cite{mitsisFunctionalBrainNetworks2020}. Future work should investigate a fuller range of signal features and seek to answer questions such as whether different pathologies display different profiles of disruption in rhythms. Nevertheless, relative band power is a primary set of features that is both widely accepted and used in neuroscience, making it easier to compare and put our results into context. We are also encouraged that our main results replicate over all frequency bands. Finally, our results are in agreement with a wider literature: diminished rhythmicity appears to be consistently observed in multiple modalities and organs in ageing and pathology \cite{kondratova2012circadian,Brooks2023}.

Another open question is if the electrophysiological circadian disruptions are a cause or consequence of the epileptogenic pathology. In our data, we observed no correlation between the resection of brain areas showing diminished rhythms and subsequent surgical outcome. This observation would suggest that sparing regions with diminished rhythms does not cause recurrent seizures after surgery \textit{per se}. Similarly, we observed no temporal correlation between the magnitude of diminished rhythms and seizure occurrence. These observations should be validated in a larger cohort in future, ideally with more complete spatial sampling. We can nevertheless conclude that we see no evidence of disrupted rhythms causing seizures directly or acutely in our cohort. This however, does not rule out indirect effects of disrupted rhythms over longer timescale, where disrupted rhythms impact sleep or other restorative processes over days and weeks and as a consequence provoke or at least enable seizures.

Interestingly, circadian and multi-day rhythms have attracted great interest recently: Various EEG and physiological signals display circadian and multi-day rhythms, and seizures occurrence\cite{karoly2021cycles,leguia2021seizure,gregg2023seizure}, as well as other seizure features\cite{schroeder2020seizure,panagiotopoulouFluctuationsEEGBand2022,gascoigne2023library} couple to the phase of these rhythms in most patients. However, the mechanism of this coupling remains elusive, and the key question is \emph{how} the seizure-generating process was modulated. Our recordings were too short for most patients to investigate multi-day rhythms, but our work can contribute towards the circadian aspect of the question. Our data suggests the epileptogenic pathology actually experiences a persistently diminished circadian modulation, most likely caused by the pathology itself. Therefore, we suggest that the circadian modulation of seizure occurrence and other seizure features is most likely mediated as a network phenomenon with other non-pathological regions, rather than arising locally within the pathological tissue itself. This supposition is also supported by recent observations that interictal markers of epilepsy are also more persistent, and less rhythmic in pathological tissue, possibly independent of seizure occurrence \cite{petito2022diurnal,wang2023temporal}. Future work will investigate if direct evidence of this network modulation can be found.

In general, a bi-directional relationship is often highlighted between circadian rhythmicity and neurological conditions\cite{Logan2019}, particularly epilepsy\cite{maganti2021untangling}. A vicious circle is often described, where pathology or pathological events may erode or disrupt circadian rhythms, and disrupted rhythms may in turn exacerbate disease symptoms or even support disease progression. A key component in this vicious circle is sleep, which we have not investigated explicitly in our study. However, future work should consider if alterations in both ultradian and circadian rhythms may reflect a loss of healthy sleep physiology regionally and behaviorally. Much of this complex interplay between rhythms, sleep, and disease is still unknown; nevertheless, our work provides initial evidence for a vital part of this vicious circle: circadian modulation of electrophysiology is indeed impaired in pathological tissue in humans \textit{in vivo}.

A limitation of our study is the coverage of electrodes in each patient. As the implantation of the iEEG is to inform surgery, electrodes are typically confined to a single hemisphere or a series of neighboring or connected regions. This means that each patient does not have complete data across all brain areas, and the AUC values calculated are based only implanted regions. A related limitation is that pathology might be present more broadly across the brain than the implanted region or SOZ \citep{horsley2023complementary,owen2023interictal,Geier2015}. It is also possible that the typical chronobiological rhythms of patients could be disrupted by surgery and implantation of the iEEG electrodes, their stay in the epilepsy monitoring unit, and changes to anti-seizure medication. 
An additional limitation of this study is that this cohort may not be representative of people with epilepsy in general because they are only those for whom it has been appropriate to undergo epilepsy surgery as they have drug-refractory focal epilepsy. A related question whether the diminished rhythms we see may be associated with specific genetic markers of epilepsy. Future work should also seek to investigate the whether specific genetic markers may help to explain the strongly diminished rhythms are seen in some individuals. 

Future work will also seek to establish a normative map \citep{groppe2013, frauscher2018,taylorNormativeBrainMapping2022a,Bernabei2022} of the expected power of each ultradian and circadian rhythms in healthy tissue. Such a map would allow a better estimation if any particular observed rhythm was abnormal in a given brain region. To create such a normative map, we need to draw on a large and diverse cohort of subjects and assess their healthy tissue to estimate the expected power. This approach has previously shown success in the context of intracranial EEG and epilepsy \citep{taylorNormativeBrainMapping2022a, Bernabei2022}, but was limited to assess short-term signal properties. Future work should extend this line of work into ultradian and circadian rhythms. 

Finally, a key limitation was our analysis of the ultradian rhythms. Due to the lack of a central pacemaker, ultradian rhythms are extremely variable between subjects, and possibly also over longer time periods. Our coarse grouping of these rhythms is inspired by literature, but not data-driven. Our results in this context should thus be interpreted as a lack of fluctuations on timescales below one day in pathological tissue, rather than any specific rhythmic activity being diminished. Future work should investigate these fluctuations from the perspective of episodic, but most likely not strictly rhythmic events \cite{goh2019episodic}.

In summary, pathological brain regions show weakened chronobiological rhythms of brain activity. At present, the causal direction of this association is not clear, but the effect is independent of seizure occurrence and brain region. These findings encourage future research to consider chronobiological rhythms in the development of disease models and treatments.

\newpage

\section{Methods}

\subsection{Preprocessing of long-term iEEG recordings}

We analysed long-term iEEG recordings from 38 subjects with refractory focal epilepsy from the National Hospital for Neurology and Neurosurgery (Table \ref{tab:tab1}). Data was stored in the National Epilepsy \& Neurology Database \citep{NENDB} and processed with approval of Newcastle University Ethics Committee (42569/2023). All subjects consented to the use of their data.
    
For each subject we processed their entire available iEEG recordings. Firstly, we divided each subject's iEEG data into 30~s non-overlapping, consecutive time segments. All channels in each time segment were re-referenced to a common average reference. In each time segment, we excluded any noisy channels (with outlier amplitude ranges) from the computed common average.
To remove power line noise, each time segment was notch filtered at $50$~Hz. Finally, segments were band-pass filtered from $0.5-80$~Hz using a $4^{\text{th}}$ order zero-phase Butterworth filter (second order forward and backward filter applied) and further downsampled to $200$~Hz. Missing data were not tolerated in any time segment and denoted as missing for the downstream analysis.

We then calculated the iEEG band power for each 30~s time segment for all channels. We extracted iEEG band power from 30~s non-overlapping iEEG segments in five frequency bands ($\delta:~1-4$~Hz, $\theta:~4-8$ Hz, $\alpha:~8-13$~Hz, $\beta:~13-30$~Hz and $\gamma:~30-47.5$~Hz, $52.5-57.5$~Hz, $62.5-77.5$~Hz) using Welch's method with 3~s non-overlapping windows. In detail, for each channel in every 2~s window we calculated the power spectral density (PSD) and used Simpson's rule to obtain the band power values which then averaged over all time windows within a 30~s segment to get the final band power values. In order to remove electrical noise, we selected custom range limits for the gamma frequency band. We $log_{10}$-transformed and normalised the band power values to sum to one for each 30~s segment. We then averaged over the relative log band power of all electrodes included in each ROI, thus obtaining, for each subject, one matrix of relative log band power at the ROI level for each frequency band (of size number of ROIs by number of 30~s segments). 

In the majority of resulting relative band power matrices had missing data, which we imputed (Suppl.~\ref{suppl:imputation}) to enable extraction of rhythms and their time-varying characteristics, such as instantaneous frequency and amplitude. Imputed data were not used for subsequent analysis, and segments with imputed data were blanked after extraction of rhythms.

\subsection{Extracting chronobiological rhythms using bandpass filter\label{methods:extract_cycles_bandpass}}

To extract rhythms of various timescales from the relative log band power time series, we performed a $4^{\text{th}}$ order zero-phase Butterworth filter (second order forward and backward). Within each subject, we extracted ultradian rhythms in different period bands (1h-3h, 3h-6h, 6h-9h, 9h-12h, 12h-9h). Finally, we denoted a circadian rhythm with period length of 19h-1.3d. Future work may consider wavelet-based methods, or empirical mode decomposition methods for this step, but for simplicity and ease of adoption, we opted for a simple filter in this study.

\subsection{MRI processing for identifying regions and resected tissue}
    
To map electrode coordinates to brain regions we used the same methods as described previously \citep{wang2023temporal}. In brief, we assigned electrodes to one of 128 regions from the Lausanne scale60 atlas \citep{hagmannMappingStructuralCore2008}. We used FreeSurfer to generate volumetric parcellations of each patient's pre-operative MRI \citep{hagmannMappingStructuralCore2008, fischlFreeSurfer2012}. Each electrode contact was assigned to the closest grey matter volumetric region within 5~mm. If the closest grey matter region was $>$5mm away then the contact was excluded from further analysis. 

To identify which regions were later resected, we used previously described methods \citep{taylorImpactEpilepsySurgery2018, taylorNormativeBrainMapping2022a}. We registered post-operative MRI to the pre-operative MRI and manually delineated the resection cavity. This manual delineation accounted for post-operative brain shift and sagging into the resection cavity. Electrode contacts within 5mm of the resection were assigned as resected. Regions with $>$25\% of their electrode contacts removed were considered as resected for downstream analysis. 

\subsection{Determining if chronobiological rhythms are diminished in pathological tissue\label{methods:auc_computation}}

In order to quantify the strength of each band power chronobiological rhythm at the ROI level for each subject, we computed the average power of each rhythm obtained from the bandpass filtered signal. In order to compare the strength of each rhythm between pathological and healthy tissue, we computed the area under the receiver operating curve (AUC). AUC values higher than 0.5 indicate diminished band power rhythms in pathological tissue, while AUCs lower than 0.5 indicate diminished band power rhythms in spared ROIs and finally AUC = 0.5 indicates no discrimination between pathological and healthy tissue. To test whether a distribution of AUC values was significantly greater than 0.5 we used a one-sided Wilcoxon ranksum test, and indeed the AUC is a normalised version of the Wilcoxon ranksum test statistic. 

We used the clinically identified regions where seizures originate (seizure onset zone, or SOZ) as a proxy for pathological tissue. To ensure robustness of our results, we also used the alternative definition of the tissue that was subsequently surgically removed, as it was deemed central to generating epileptic seizures. We obtained similar results (Suppl.~\ref{suppl:resvsspared}) with either definition, and will use the broad term of ``pathological tissue'' throughout the paper to refer to tissue that is most likely functionally (causing seizures) and/or structurally pathological.

\subsection{Fitting a mixed effects model to control for brain area}
\label{sec:model}
To show that pathological tissue predicted diminished rhythms of brain activity above any brain area specific effect, we fit a mixed effects model predicting the rhythm power in a given ROI from a categorical variable representing lobe within which the ROI resides, and whether that ROI was pathological, grouping by a random patient offset. Equation~\ref{memod1} describes the model, where $RhythmPower_{i,j}$ is the power of the rhythm for patient $i$ in ROI $j$, $\beta_0$ is the fixed intercept, $\beta_1$ is the fixed effect for pathological ROI, $Path_{i,j}$ is a binary variable indicating that ROI $j$ of patient $i$ is pathological, $\beta_{2..7}$ is the fixed effect of the region being in each lobe - a categorical variable with 7 levels corresponding to the x axis of figure 3 (a), $Lobe_{i,j}$ indicates whether ROI $j$ of patient $i$ is in each lobe, $u_i$ is the random intercept for patient $i$, and $\mathcal{E}_{i,j}$ is the error. Regions were grouped into 7 possible lobes/areas: frontal, temporal, parietal, occipital, cingulate, hippocampus, amygdala. The dependent variable (rhythm power) was standardized before fitting to give partially standardized coefficients, allowing us to interpret $\beta_{1}$ as the expected effect of pathology on rhythm power (in standard deviations) \citep{Lorah2018}. A likelihood ratio test was used to assess whether the effect of pathology was statistically significant - comparing to a reduced model without pathology as a variable. Maximum likelihood estimation was used in the fitting of models, with Python (3.8.10) alongside the statsmodels library (0.14.0) and its MixedLM function providing the software.

\begin{equation} \label{memod1}
rhythmPower_{i,j} = \beta_0 + \beta_1 hasPath_{i,j} + \beta_{2..7} Lobe_{i,j} + u_i + \mathcal{E}_{i,j}
\end{equation}

\subsection{Data availability}
Data used is available at: \url{https://zenodo.org/record/8289342}

\subsection{Code availability}
All analysis was performed using Python (version 3.8) and MATLAB (version R2023a). \\
Code used is available at: \url{https://github.com/cnnp-lab/DiminishedRhythmsPathology} \\

\subsection{Author Contributions Statement}
\begin{itemize}
  \item Conceptualization: MP PNT YW
  \item Methodology: MP CT BCS YW
  \item Software/validation: CT MP YW NE
  \item Formal analysis: CT MP YW
  \item Resources: FC BD JSD AMc AM 
  \item Data curation: MP CT SJG GB YW FC BD JSD AMc AM 
  \item Writing: CT MP PNT YW
  \item Supervision: PNT YW
\end{itemize}
\subsection{Competing Interests Statement}
The authors declare no competing interests.

\newpage
\bibliography{ref}

\begin{thebibliography}{51}
\providecommand{\natexlab}[1]{#1}
\providecommand{\url}[1]{\texttt{#1}}
\expandafter\ifx\csname urlstyle\endcsname\relax
  \providecommand{\doi}[1]{doi: #1}\else
  \providecommand{\doi}{doi: \begingroup \urlstyle{rm}\Url}\fi

\bibitem[NEN()]{NENDB}
The national epilepsy and neurology database.
\newblock \url{https://www.hra.nhs.uk/planning-and-improving-research/application-summaries/research-summaries/national-epilepsy-neurology-database/}.
\newblock Accessed: 2024-02-07.

\bibitem[Aeschbach et~al.(1999)Aeschbach, Matthews, Postolache, Jackson, Giesen, and Wehr]{aeschbach1999two}
Daniel Aeschbach, Jeffery~R Matthews, Teodor~T Postolache, Michael~A Jackson, Holly~A Giesen, and Thomas~A Wehr.
\newblock Two circadian rhythms in the human electroencephalogram during wakefulness.
\newblock \emph{American Journal of Physiology-Regulatory, Integrative and Comparative Physiology}, 277\penalty0 (6):\penalty0 R1771--R1779, 1999.

\bibitem[Basu et~al.(2021)Basu, Maguire, and Salpekar]{Basu2021}
Trina Basu, Jamie Maguire, and Jay~A Salpekar.
\newblock Hypothalamic-pituitary-adrenal axis targets for the treatment of epilepsy.
\newblock \emph{Neuroscience letters}, 746:\penalty0 135618, 2 2021.
\newblock ISSN 1872-7972.
\newblock \doi{10.1016/j.neulet.2020.135618}.

\bibitem[Bernabei et~al.(2022)Bernabei, Sinha, Arnold, Conrad, Ong, Pattnaik, Stein, Shinohara, Lucas, Bassett, Davis, and Litt]{Bernabei2022}
John~M Bernabei, Nishant Sinha, T~Campbell Arnold, Erin Conrad, Ian Ong, Akash~R Pattnaik, Joel~M Stein, Russell~T Shinohara, Timothy~H Lucas, Dani~S Bassett, Kathryn~A Davis, and Brian Litt.
\newblock Normative intracranial eeg maps epileptogenic tissues in focal epilepsy.
\newblock \emph{Brain : a journal of neurology}, 145:\penalty0 1949--1961, 6 2022.
\newblock ISSN 1460-2156.
\newblock \doi{10.1093/brain/awab480}.
\newblock URL \url{http://www.ncbi.nlm.nih.gov/pubmed/35640886 http://www.pubmedcentral.nih.gov/articlerender.fcgi?artid=PMC9630716}.

\bibitem[Brooks et~al.(2023)Brooks, Lahens, Sheline, Fitzgerald, and Skarke]{Brooks2023}
Thomas Brooks, Nicholas Lahens, Yvette Sheline, Garret Fitzgerald, and Carsten Skarke.
\newblock Diurnal rhythmicity of wearable device-measured wrist temperature predicts future disease incidence in the uk biobank.
\newblock \emph{Research Square. Preprint.}, 2 2023.
\newblock \doi{10.21203/RS.3.RS-2535978/V1}.

\bibitem[Chan and Liu(2021)]{chan2021}
Felix Chan and Judy Liu.
\newblock Molecular regulation of brain metabolism underlying circadian epilepsy.
\newblock \emph{Epilepsia}, 62:\penalty0 S32--S48, 2021.

\bibitem[Croce et~al.(2018)Croce, Quercia, Costa, and Zappasodi]{croce2018circadian}
Pierpaolo Croce, Angelica Quercia, Sergio Costa, and Filippo Zappasodi.
\newblock Circadian rhythms in fractal features of eeg signals.
\newblock \emph{Frontiers in physiology}, 9:\penalty0 1567, 2018.

\bibitem[Fawcett(2006)]{Fawcett2006}
Tom Fawcett.
\newblock An introduction to roc analysis.
\newblock \emph{Pattern Recognition Letters}, 27:\penalty0 861--874, 6 2006.
\newblock ISSN 01678655.
\newblock \doi{10.1016/j.patrec.2005.10.010}.
\newblock URL \url{https://linkinghub.elsevier.com/retrieve/pii/S016786550500303X}.

\bibitem[Fischl(2012)]{fischlFreeSurfer2012}
Bruce Fischl.
\newblock {{FreeSurfer}}.
\newblock \emph{NeuroImage}, 62\penalty0 (2):\penalty0 774--781, August 2012.
\newblock ISSN 10538119.
\newblock \doi{10.1016/j.neuroimage.2012.01.021}.
\newblock URL \url{https://linkinghub.elsevier.com/retrieve/pii/S1053811912000389}.

\bibitem[Frauscher et~al.(2018)Frauscher, Von~Ellenrieder, Zelmann, Dole{\v{z}}alov{\'a}, Minotti, Olivier, Hall, Hoffmann, Nguyen, Kahane, et~al.]{frauscher2018}
Birgit Frauscher, Nicolas Von~Ellenrieder, Rina Zelmann, Irena Dole{\v{z}}alov{\'a}, Lorella Minotti, Andre Olivier, Jeffery Hall, Dominique Hoffmann, Dang~Khoa Nguyen, Philippe Kahane, et~al.
\newblock Atlas of the normal intracranial electroencephalogram: neurophysiological awake activity in different cortical areas.
\newblock \emph{Brain}, 141\penalty0 (4):\penalty0 1130--1144, 2018.

\bibitem[Gascoigne et~al.(2023)Gascoigne, Waldmann, Schroeder, Panagiotopoulou, Blickwedel, Chowdhury, Cronie, Diehl, Duncan, Falconer, et~al.]{gascoigne2023library}
Sarah~J Gascoigne, Leonard Waldmann, Gabrielle~M Schroeder, Mariella Panagiotopoulou, Jess Blickwedel, Fahmida Chowdhury, Alison Cronie, Beate Diehl, John~S Duncan, Jennifer Falconer, et~al.
\newblock A library of quantitative markers of seizure severity.
\newblock \emph{Epilepsia}, 2023.

\bibitem[Geier et~al.(2015)Geier, Bialonski, Elger, and Lehnertz]{Geier2015}
Christian Geier, Stephan Bialonski, Christian~E Elger, and Klaus Lehnertz.
\newblock How important is the seizure onset zone for seizure dynamics?
\newblock \emph{Seizure}, 25:\penalty0 160--6, 2 2015.
\newblock ISSN 1532-2688.
\newblock \doi{10.1016/j.seizure.2014.10.013}.
\newblock URL \url{http://www.ncbi.nlm.nih.gov/pubmed/25468511}.

\bibitem[George et~al.(2017)George, Birnie, Flynn, Kershaw, Lightman, and Conway-Campbell]{George2017}
Charlotte~L George, Matthew~T Birnie, Benjamin~P Flynn, Yvonne~M Kershaw, Stafford~L Lightman, and Becky~L Conway-Campbell.
\newblock Ultradian glucocorticoid exposure directs gene-dependent and tissue-specific mrna expression patterns in vivo.
\newblock \emph{Molecular and cellular endocrinology}, 439:\penalty0 46--53, 1 2017.
\newblock ISSN 1872-8057.
\newblock \doi{10.1016/j.mce.2016.10.019}.
\newblock URL \url{http://www.ncbi.nlm.nih.gov/pubmed/27769714 http://www.pubmedcentral.nih.gov/articlerender.fcgi?artid=PMC5131830}.

\bibitem[Goh et~al.(2019)Goh, Maloney, Mark, and Blache]{goh2019episodic}
Grace~H Goh, Shane~K Maloney, Peter~J Mark, and Dominique Blache.
\newblock Episodic ultradian events—ultradian rhythms.
\newblock \emph{Biology}, 8\penalty0 (1):\penalty0 15, 2019.

\bibitem[Gregg et~al.(2023)Gregg, Pal~Attia, Nasseri, Joseph, Karoly, Cui, Stirling, Viana, Richner, Nurse, et~al.]{gregg2023seizure}
Nicholas~M Gregg, Tal Pal~Attia, Mona Nasseri, Boney Joseph, Philippa Karoly, Jie Cui, Rachel~E Stirling, Pedro~F Viana, Thomas~J Richner, Ewan~S Nurse, et~al.
\newblock Seizure occurrence is linked to multiday cycles in diverse physiological signals.
\newblock \emph{Epilepsia}, 64\penalty0 (6):\penalty0 1627--1639, 2023.

\bibitem[Groppe et~al.(2013)Groppe, Bickel, Keller, Jain, Hwang, Harden, and Mehta]{groppe2013}
David~M Groppe, Stephan Bickel, Corey~J Keller, Sanjay~K Jain, Sean~T Hwang, Cynthia Harden, and Ashesh~D Mehta.
\newblock Dominant frequencies of resting human brain activity as measured by the electrocorticogram.
\newblock \emph{Neuroimage}, 79:\penalty0 223--233, 2013.

\bibitem[Hagmann et~al.(2008)Hagmann, Cammoun, Gigandet, Meuli, Honey, Wedeen, and Sporns]{hagmannMappingStructuralCore2008}
Patric Hagmann, Leila Cammoun, Xavier Gigandet, Reto Meuli, Christopher~J Honey, Van~J Wedeen, and Olaf Sporns.
\newblock Mapping the {{Structural Core}} of {{Human Cerebral Cortex}}.
\newblock \emph{PLoS Biology}, 6\penalty0 (7):\penalty0 e159, July 2008.
\newblock ISSN 1545-7885.
\newblock \doi{10.1371/journal.pbio.0060159}.
\newblock URL \url{https://dx.plos.org/10.1371/journal.pbio.0060159}.

\bibitem[Hartmann et~al.(1997)Hartmann, Veldhuis, Deuschle, Standhardt, and Heuser]{Hartmann1997}
A.~Hartmann, J.~D. Veldhuis, M.~Deuschle, H.~Standhardt, and I.~Heuser.
\newblock Twenty-four hour cortisol release profiles in patients with alzheimer’s and parkinson’s disease compared to normal controls: Ultradian secretory pulsatility and diurnal variation.
\newblock \emph{Neurobiology of Aging}, 18:\penalty0 285--289, 5 1997.
\newblock ISSN 0197-4580.
\newblock \doi{10.1016/S0197-4580(97)80309-0}.

\bibitem[Henley et~al.(2009)Henley, Leendertz, Russell, Wood, Taheri, Woltersdorf, and Lightman]{Henley2009}
D~E Henley, J~A Leendertz, G~M Russell, S~A Wood, S~Taheri, W~W Woltersdorf, and S~L Lightman.
\newblock Development of an automated blood sampling system for use in humans.
\newblock \emph{Journal of medical engineering \& technology}, 33:\penalty0 199--208, 4 2009.
\newblock ISSN 1464-522X.
\newblock \doi{10.1080/03091900802185970}.
\newblock URL \url{http://www.ncbi.nlm.nih.gov/pubmed/19340690}.

\bibitem[Horsley et~al.(2023)Horsley, Thomas, Chowdhury, Diehl, McEvoy, Miserocchi, de~Tisi, Vos, Walker, Winston, et~al.]{horsley2023complementary}
Jonathan~J Horsley, Rhys~H Thomas, Fahmida~A Chowdhury, Beate Diehl, Andrew~W McEvoy, Anna Miserocchi, Jane de~Tisi, Skoerd~B Vos, Matthew~C Walker, Gavin~P Winston, et~al.
\newblock Complementary structural and functional abnormalities to localise epileptogenic tissue.
\newblock \emph{arXiv preprint arXiv:2304.03192}, 2023.

\bibitem[Hughey and Butte(2016)]{Hughey2016}
Jacob~J Hughey and Atul~J Butte.
\newblock Differential phasing between circadian clocks in the brain and peripheral organs in humans.
\newblock \emph{Journal of biological rhythms}, 31:\penalty0 588--597, 12 2016.
\newblock ISSN 1552-4531.
\newblock \doi{10.1177/0748730416668049}.

\bibitem[Karoly et~al.(2021)Karoly, Rao, Gregg, Worrell, Bernard, Cook, and Baud]{karoly2021cycles}
Philippa~J Karoly, Vikram~R Rao, Nicholas~M Gregg, Gregory~A Worrell, Christophe Bernard, Mark~J Cook, and Maxime~O Baud.
\newblock Cycles in epilepsy.
\newblock \emph{Nature Reviews Neurology}, 17\penalty0 (5):\penalty0 267--284, 2021.

\bibitem[Kondratova and Kondratov(2012)]{kondratova2012circadian}
Anna~A Kondratova and Roman~V Kondratov.
\newblock The circadian clock and pathology of the ageing brain.
\newblock \emph{Nature Reviews Neuroscience}, 13\penalty0 (5):\penalty0 325--335, 2012.

\bibitem[Leguia et~al.(2021)Leguia, Andrzejak, Rummel, Fan, Mirro, Tcheng, Rao, and Baud]{leguia2021seizure}
Marc~G Leguia, Ralph~G Andrzejak, Christian Rummel, Joline~M Fan, Emily~A Mirro, Thomas~K Tcheng, Vikram~R Rao, and Maxime~O Baud.
\newblock Seizure cycles in focal epilepsy.
\newblock \emph{JAMA neurology}, 78\penalty0 (4):\penalty0 454--463, 2021.

\bibitem[Leise(2015)]{Leise2015}
Tanya~L Leise.
\newblock Wavelet-based analysis of circadian behavioral rhythms.
\newblock \emph{Methods in enzymology}, 551:\penalty0 95--119, 1 2015.

\bibitem[Logan and McClung(2019)]{Logan2019}
Ryan~W Logan and Colleen~A McClung.
\newblock Rhythms of life: circadian disruption and brain disorders across the lifespan.
\newblock \emph{Nature reviews. Neuroscience}, 20:\penalty0 49--65, 1 2019.
\newblock ISSN 1471-0048.
\newblock \doi{10.1038/s41583-018-0088-y}.
\newblock URL \url{http://www.ncbi.nlm.nih.gov/pubmed/30459365 http://www.pubmedcentral.nih.gov/articlerender.fcgi?artid=PMC6338075}.

\bibitem[Lorah(2018)]{Lorah2018}
Julie Lorah.
\newblock Effect size measures for multilevel models: definition, interpretation, and timss example.
\newblock \emph{Large-scale Assessments in Education}, 6:\penalty0 8, 12 2018.
\newblock ISSN 2196-0739.
\newblock \doi{10.1186/s40536-018-0061-2}.
\newblock URL \url{https://largescaleassessmentsineducation.springeropen.com/articles/10.1186/s40536-018-0061-2}.

\bibitem[Maganti and Jones(2021)]{maganti2021untangling}
Rama~K Maganti and Mathew~V Jones.
\newblock Untangling a web: basic mechanisms of the complex interactions between sleep, circadian rhythms, and epilepsy.
\newblock \emph{Epilepsy Currents}, 21\penalty0 (2):\penalty0 105--110, 2021.

\bibitem[Mitsis et~al.(2020)Mitsis, Anastasiadou, Christodoulakis, Papathanasiou, Papacostas, and Hadjipapas]{mitsisFunctionalBrainNetworks2020}
Georgios~D. Mitsis, Maria~N. Anastasiadou, Manolis Christodoulakis, Eleftherios~S. Papathanasiou, Savvas~S. Papacostas, and Avgis Hadjipapas.
\newblock Functional brain networks of patients with epilepsy exhibit pronounced multiscale periodicities, which correlate with seizure onset.
\newblock \emph{Human Brain Mapping}, 41\penalty0 (8):\penalty0 2059--2076, 2020.
\newblock ISSN 1097-0193.
\newblock \doi{10.1002/hbm.24930}.
\newblock URL \url{https://onlinelibrary.wiley.com/doi/abs/10.1002/hbm.24930}.

\bibitem[Monje et~al.(2017)Monje, Cicvaric, Aguilar, Elbau, Horvath, Diao, Glat, and Pollak]{Monje2017}
Francisco~J. Monje, Ana Cicvaric, Juan Pablo~Acevedo Aguilar, Immanuel Elbau, Orsolya Horvath, Weifei Diao, Micaela Glat, and Daniela~D. Pollak.
\newblock Disrupted ultradian activity rhythms and differential expression of several clock genes in interleukin-6-deficient mice.
\newblock \emph{Frontiers in Neurology}, 8:\penalty0 22, 3 2017.
\newblock ISSN 16642295.
\newblock \doi{10.3389/FNEUR.2017.00099/FULL}.

\bibitem[Mure et~al.(2018)Mure, Le, Benegiamo, Chang, Rios, Jillani, Ngotho, Kariuki, Dkhissi-Benyahya, Cooper, et~al.]{mure2018}
Ludovic~S Mure, Hiep~D Le, Giorgia Benegiamo, Max~W Chang, Luis Rios, Ngalla Jillani, Maina Ngotho, Thomas Kariuki, Ouria Dkhissi-Benyahya, Howard~M Cooper, et~al.
\newblock Diurnal transcriptome atlas of a primate across major neural and peripheral tissues.
\newblock \emph{Science}, 359\penalty0 (6381):\penalty0 eaao0318, 2018.

\bibitem[Owen et~al.(2023)Owen, Janiukstyte, Hall, Chowdhury, Diehl, McEvoy, Miserocchi, de~Tisi, Duncan, Rugg-Gunn, et~al.]{owen2023interictal}
Thomas~W Owen, Vytene Janiukstyte, Gerard~R Hall, Fahmida~A Chowdhury, Beate Diehl, Andrew McEvoy, Anna Miserocchi, Jane de~Tisi, John~S Duncan, Fergus Rugg-Gunn, et~al.
\newblock Interictal meg abnormalities to guide intracranial electrode implantation and predict surgical outcome.
\newblock \emph{arXiv preprint arXiv:2304.05199}, 2023.

\bibitem[Panagiotopoulou et~al.(2022)Panagiotopoulou, Papasavvas, Schroeder, Thomas, Taylor, and Wang]{panagiotopoulouFluctuationsEEGBand2022}
M.~Panagiotopoulou, C.~A. Papasavvas, G.~M. Schroeder, R.~H. Thomas, P.~N. Taylor, and Y.~Wang.
\newblock Fluctuations in {{EEG}} band power at subject-specific timescales over minutes to days explain changes in seizure evolutions.
\newblock \emph{Human Brain Mapping}, 2022.
\newblock \doi{10.1002/hbm.25796}.

\bibitem[Panda et~al.(2002)Panda, Antoch, Miller, Su, Schook, Straume, Schultz, Kay, Takahashi, and Hogenesch]{panda2002}
Satchidananda Panda, Marina~P Antoch, Brooke~H Miller, Andrew~I Su, Andrew~B Schook, Marty Straume, Peter~G Schultz, Steve~A Kay, Joseph~S Takahashi, and John~B Hogenesch.
\newblock Coordinated transcription of key pathways in the mouse by the circadian clock.
\newblock \emph{Cell}, 109\penalty0 (3):\penalty0 307--320, 2002.

\bibitem[Petito et~al.(2022)Petito, Housekeeper, Buroker, Scholle, Ervin, Frink, Greiner, Skoch, Mangano, Dye, et~al.]{petito2022diurnal}
Gabrielle~T Petito, Jeremy Housekeeper, Jason Buroker, Craig Scholle, Brian Ervin, Clayton Frink, Hansel~M Greiner, Jesse Skoch, Francesco~T Mangano, Thomas~J Dye, et~al.
\newblock Diurnal rhythms of spontaneous intracranial high-frequency oscillations.
\newblock \emph{Seizure}, 102:\penalty0 105--112, 2022.

\bibitem[Philippu(2016)]{Philippu2016}
Athineos Philippu.
\newblock Nitric oxide: A universal modulator of brain function.
\newblock \emph{Current medicinal chemistry}, 23:\penalty0 2643--2652, 9 2016.
\newblock ISSN 1875-533X.
\newblock \doi{10.2174/0929867323666160627120408}.
\newblock URL \url{https://pubmed.ncbi.nlm.nih.gov/27356532/}.

\bibitem[Philippu(2019)]{Philippu2019}
Athineos Philippu.
\newblock Neurotransmitters are released in brain areas according to ultradian rhythms: Coincidence with ultradian oscillations of eeg waves.
\newblock \emph{Journal of chemical neuroanatomy}, 96:\penalty0 66--72, 3 2019.
\newblock ISSN 1873-6300.
\newblock \doi{10.1016/J.JCHEMNEU.2018.12.007}.
\newblock URL \url{https://pubmed.ncbi.nlm.nih.gov/30576780/}.

\bibitem[Polich(1997)]{Polich1997}
J.~Polich.
\newblock On the relationship between eeg and p300: individual differences, aging, and ultradian rhythms.
\newblock \emph{International journal of psychophysiology : official journal of the International Organization of Psychophysiology}, 26:\penalty0 299--317, 6 1997.
\newblock ISSN 0167-8760.
\newblock \doi{10.1016/S0167-8760(97)00772-1}.
\newblock URL \url{https://pubmed.ncbi.nlm.nih.gov/9203011/}.

\bibitem[Polich and Kok(1995)]{Polich1995}
John Polich and Albert Kok.
\newblock Cognitive and biological determinants of p300: an integrative review.
\newblock \emph{Biological Psychology}, 41:\penalty0 103--146, 1995.
\newblock ISSN 03010511.
\newblock \doi{10.1016/0301-0511(95)05130-9}.
\newblock URL \url{https://pubmed.ncbi.nlm.nih.gov/8534788/}.

\bibitem[Ravden and Polich(1999)]{Ravden1999}
Daran Ravden and John Polich.
\newblock On p300 measurement stability: habituation, intra-trial block variation, and ultradian rhythms.
\newblock \emph{Biological psychology}, 51:\penalty0 59--76, 1999.
\newblock ISSN 0301-0511.
\newblock \doi{10.1016/S0301-0511(99)00015-0}.
\newblock URL \url{https://pubmed.ncbi.nlm.nih.gov/10579421/}.

\bibitem[Refinetti et~al.(2007)Refinetti, Cornélissen, and Halberg]{Refinetti2007}
Roberto Refinetti, Germaine Cornélissen, and Franz Halberg.
\newblock Procedures for numerical analysis of circadian rhythms.
\newblock \emph{Biological Rhythm Research}, 38:\penalty0 275--325, 8 2007.

\bibitem[Rozhkova et~al.(2021)Rozhkova, Teply, and Bazhanova]{Rozhkova2021}
I.~S. Rozhkova, D.~L. Teply, and E.~D. Bazhanova.
\newblock Ultradian rhythms and oxidative stress in lymph-node tissue during ontogenesis.
\newblock \emph{Advances in Gerontology}, 11:\penalty0 268--273, 7 2021.
\newblock ISSN 2079-0570.
\newblock \doi{10.1134/S2079057021030140}.

\bibitem[Schroeder et~al.(2020)Schroeder, Diehl, Chowdhury, Duncan, de~Tisi, Trevelyan, Forsyth, Jackson, Taylor, and Wang]{schroeder2020seizure}
Gabrielle~M Schroeder, Beate Diehl, Fahmida~A Chowdhury, John~S Duncan, Jane de~Tisi, Andrew~J Trevelyan, Rob Forsyth, Andrew Jackson, Peter~N Taylor, and Yujiang Wang.
\newblock Seizure pathways change on circadian and slower timescales in individual patients with focal epilepsy.
\newblock \emph{Proceedings of the National Academy of Sciences}, 117\penalty0 (20):\penalty0 11048--11058, 2020.

\bibitem[Schroeder et~al.(2023)Schroeder, Karoly, Maturana, Panagiotopoulou, Taylor, Cook, and Wang]{Schroeder2023}
Gabrielle~M. Schroeder, Philippa~J. Karoly, Matias Maturana, Mariella Panagiotopoulou, Peter~N. Taylor, Mark~J. Cook, and Yujiang Wang.
\newblock Chronic intracranial eeg recordings and interictal spike rate reveal multiscale temporal modulations in seizure states.
\newblock \emph{Brain Communications}, 5, 8 2023.

\bibitem[Tan et~al.(2003{\natexlab{a}})Tan, Uchiyama, Shibui, Tagaya, Suzuki, Kamei, Kim, Aritaka, Ozaki, and Takahashi]{tan2003circadian}
Xin Tan, Makoto Uchiyama, Kayo Shibui, Hirokuni Tagaya, Hiroyuki Suzuki, Yuichi Kamei, Kyuja Kim, Sayaka Aritaka, Akiko Ozaki, and Kiyohisa Takahashi.
\newblock Circadian rhythms in humans’ delta sleep electroencephalogram.
\newblock \emph{Neuroscience letters}, 344\penalty0 (3):\penalty0 205--208, 2003{\natexlab{a}}.

\bibitem[Tan et~al.(2003{\natexlab{b}})Tan, Uchiyama, Shibui, Tagaya, Suzuki, Kamei, Kim, Aritake, Ozaki, and Takahashi]{Tan2003}
Xin Tan, Makoto Uchiyama, Kayo Shibui, Hirokuni Tagaya, Hiroyuki Suzuki, Yuichi Kamei, Kyuja Kim, Sayaka Aritake, Akiko Ozaki, and Kiyohisa Takahashi.
\newblock Circadian rhythms in humans' delta sleep electroencephalogram.
\newblock \emph{Neuroscience letters}, 344:\penalty0 205--8, 7 2003{\natexlab{b}}.

\bibitem[Taylor et~al.(2018)Taylor, Sinha, Wang, Vos, {de Tisi}, Miserocchi, McEvoy, Winston, and Duncan]{taylorImpactEpilepsySurgery2018}
Peter~N. Taylor, Nishant Sinha, Yujiang Wang, Sjoerd~B. Vos, Jane {de Tisi}, Anna Miserocchi, Andrew~W. McEvoy, Gavin~P. Winston, and John~S. Duncan.
\newblock The impact of epilepsy surgery on the structural connectome and its relation to outcome.
\newblock \emph{NeuroImage: Clinical}, 18:\penalty0 202--214, 2018.
\newblock ISSN 22131582.
\newblock \doi{10.1016/j.nicl.2018.01.028}.
\newblock URL \url{https://linkinghub.elsevier.com/retrieve/pii/S2213158218300287}.

\bibitem[Taylor et~al.(2022)Taylor, Papasavvas, Owen, Schroeder, Hutchings, Chowdhury, Diehl, Duncan, McEvoy, Miserocchi, {de Tisi}, Vos, Walker, and Wang]{taylorNormativeBrainMapping2022a}
Peter~N Taylor, Christoforos~A Papasavvas, Thomas~W Owen, Gabrielle~M Schroeder, Frances~E Hutchings, Fahmida~A Chowdhury, Beate Diehl, John~S Duncan, Andrew~W McEvoy, Anna Miserocchi, Jane {de Tisi}, Sjoerd~B Vos, Matthew~C Walker, and Yujiang Wang.
\newblock Normative brain mapping of interictal intracranial {{EEG}} to localize epileptogenic tissue.
\newblock \emph{Brain}, 145\penalty0 (3):\penalty0 939--949, April 2022.
\newblock ISSN 0006-8950, 1460-2156.
\newblock \doi{10.1093/brain/awab380}.
\newblock URL \url{https://academic.oup.com/brain/article/145/3/939/6514463}.

\bibitem[Wang et~al.(2023)Wang, Schroeder, Horsley, Panagiotopoulou, Chowdhury, Diehl, Duncan, McEvoy, Miserocchi, de~Tisi, et~al.]{wang2023temporal}
Yujiang Wang, Gabrielle~M Schroeder, Jonathan~J Horsley, Mariella Panagiotopoulou, Fahmida~A Chowdhury, Beate Diehl, John~S Duncan, Andrew~W McEvoy, Anna Miserocchi, Jane de~Tisi, et~al.
\newblock Temporal stability of intracranial eeg abnormality maps for localizing epileptogenic tissue.
\newblock \emph{Epilepsia}, 2023.

\bibitem[Wu et~al.(2021)Wu, Liu, Liu, Meng, Du, Li, Dong, Zhang, Li, and Zhang]{Wu2021}
Hao Wu, Yong Liu, Lishuo Liu, Qiang Meng, Changwang Du, Kuo Li, Shan Dong, Yong Zhang, Huanfa Li, and Hua Zhang.
\newblock Decreased expression of the clock gene bmal1 is involved in the pathogenesis of temporal lobe epilepsy.
\newblock \emph{Molecular brain}, 14:\penalty0 113, 7 2021.
\newblock ISSN 1756-6606.
\newblock \doi{10.1186/s13041-021-00824-4}.
\newblock URL \url{http://www.ncbi.nlm.nih.gov/pubmed/34261484 http://www.pubmedcentral.nih.gov/articlerender.fcgi?artid=PMC8281660}.

\bibitem[Yeung et~al.(2018)Yeung, Mermet, Jouffe, Marquis, Charpagne, Gachon, and Naef]{Yeung2018}
Jake Yeung, Jérôme Mermet, Céline Jouffe, Julien Marquis, Aline Charpagne, Frédéric Gachon, and Felix Naef.
\newblock Transcription factor activity rhythms and tissue-specific chromatin interactions explain circadian gene expression across organs.
\newblock \emph{Genome research}, 28:\penalty0 182--191, 2 2018.
\newblock ISSN 1549-5469.
\newblock \doi{10.1101/gr.222430.117}.
\newblock URL \url{http://www.ncbi.nlm.nih.gov/pubmed/29254942 http://www.pubmedcentral.nih.gov/articlerender.fcgi?artid=PMC5793782}.

\end{thebibliography}
\newpage

\section{Acknowledgements}
We thank members of the Computational Neurology, Neuroscience \& Psychiatry Lab (www.cnnp-lab.com) for discussions on the analysis and manuscript; P.N.T. and Y.W. are both supported by UKRI Future Leaders Fellowships (MR/T04294X/1, MR/V026569/1). JSD, JdT are supported by the NIHR UCLH/UCL Biomedical Research Centre.
\section{Tables}
\begin{table}[h]
\caption{Table 1: Summary of patient data used in the analysis.}
\label{tab:tab1}
\begin{tabular}{llll}
\textbf{N}                    & 38        \\
\textbf{Age (mean,SD)}             & 32.1 (8.5) \\
\textbf{Sex (M,F)}                 & 17,21         \\
\textbf{Temporal, extratemporal}    & 17,21          \\
\textbf{Side (Left, Right)}        & 22,16        \\
\textbf{Num contacts (mean, sd)}   & 72.8 (25.2)     \\
\textbf{Recording Duration in hours (mean, sd)} & 111.5 (58.1
)    
\end{tabular}
\end{table}
\section{Figure Captions}
\begin{figure}[p]
    \centering
    \includegraphics[width=1\linewidth]{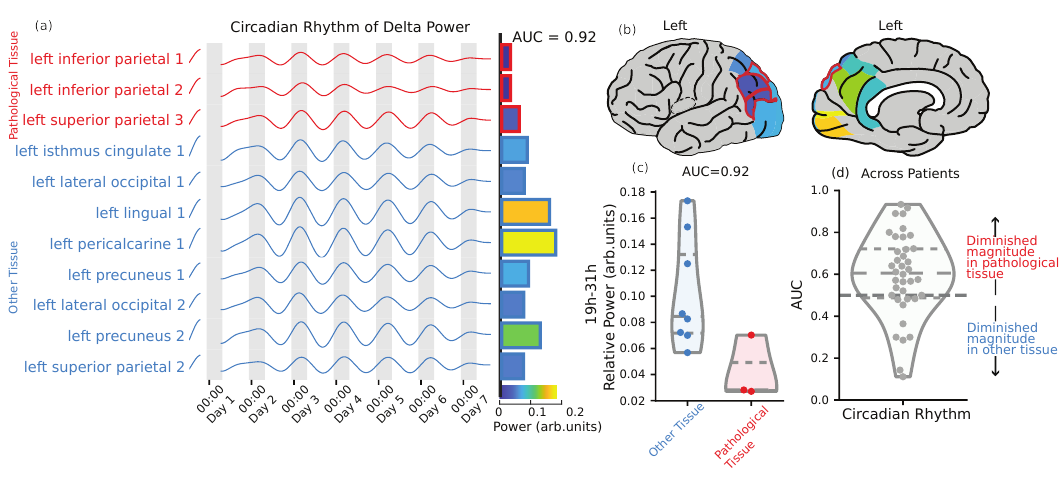}
    \caption{\textbf{Circadian rhythm of delta power is diminished in brain areas with pathology.} a-c) From subject A: (a) The circadian rhythm of delta power, obtained by applying a band-pass filter to isolate the power of the signal with a period between 19 and 31 hours. Pathological tissue is shown in red. Bars indicate the relative power of the signal in each ROI. (b) The relative power in each brain area, pathological tissue is outlined in red. (c) The relative power of the circadian rhythm in pathological tissue compared to other tissue. An AUC of 0.92 is calculated when using this relative power to predict pathology.  (d) The AUC of each subject for the same prediction. An AUC$>$0.5 indicates that the rhythm is diminished in pathological tissue. Dashed lines indicate median and quartiles of the distribution.}
    \label{fig:fig1}
\end{figure}

\begin{figure}[p]
    \centering
    \includegraphics[width=1\linewidth]{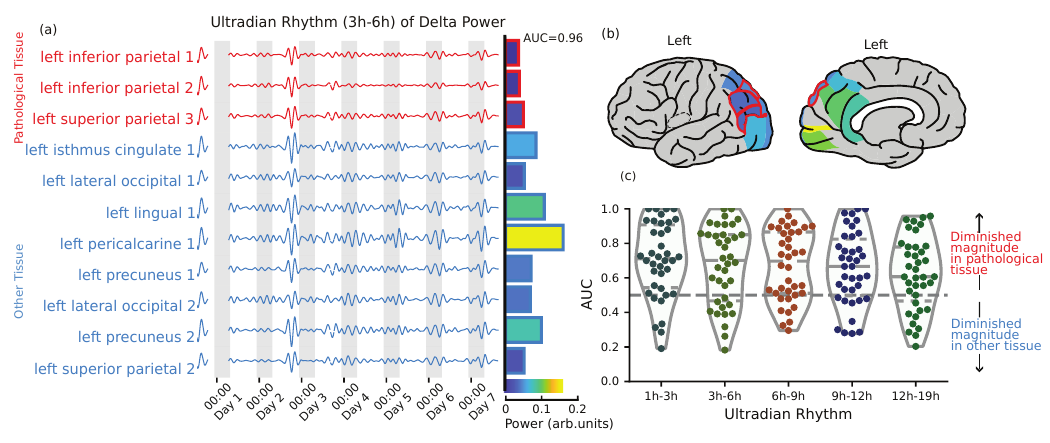}
    \caption{\textbf{Ultradian rhythm of delta power is diminished in brain areas with pathology.} a-c) From subject A: (a) An ultradian rhythm of delta power, obtained by applying a band-pass filter to isolate the power of the signal with a period between 3 and 6 hours. Pathological tissue is shown in red. Bars indicate the relative power of the signal in each ROI. (b) The relative power of the ultradian rhythm (3-6h) in each brain area, pathological tissue is outlined in red, no recording was performed in gray areas. (c) The AUC of each subject when predicting pathology from the power of each ultradian rhythm. An AUC$>$0.5 indicates that the rhythm is diminished in pathological tissue. Dashed lines indicate median and quartiles of the distribution.}
    \label{fig:fig2}
\end{figure}

\begin{figure}[p]
    \centering
    \includegraphics[width=1\linewidth]{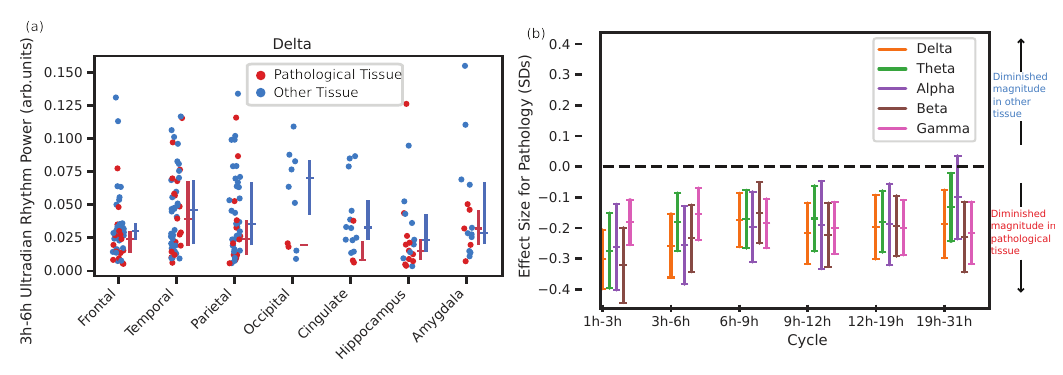}
    \caption{\textbf{Tissue pathology is associated with diminished rhythms while controlling for brain region.} (a) The power of the delta ultradian rhythm (3h-6h) in each lobe or region for all patients, with each dot showing the power of the rhythm recorded in that region - some patients have multiple electrode per region. Red dots indicate electrodes recording from pathological tissue, blue from other tissue. The lines indicate median and inter-quartile range. These are calculated by first calculating a median power for each individual in each region (if they have more than one electrode in that region) and then we calculate a population median and interquartile range across all patients who have electrodes within the region. (b) The effect size (partially standardized coefficient and 95 \% confidence interval) for tissue pathology in a mixed effects regression relating brain region and tissue pathology to rhythm power. Shows the expected difference in rhythm power (in standard deviations) for pathological tissue, while accounting for variability across regions and individuals. The centre line indicates the estimated effect size, the error bars indicate the 95\% confidence interval.}
    \label{fig:fig3}
\end{figure}

\begin{figure}[p]
    \centering
    \includegraphics[width=1\linewidth]{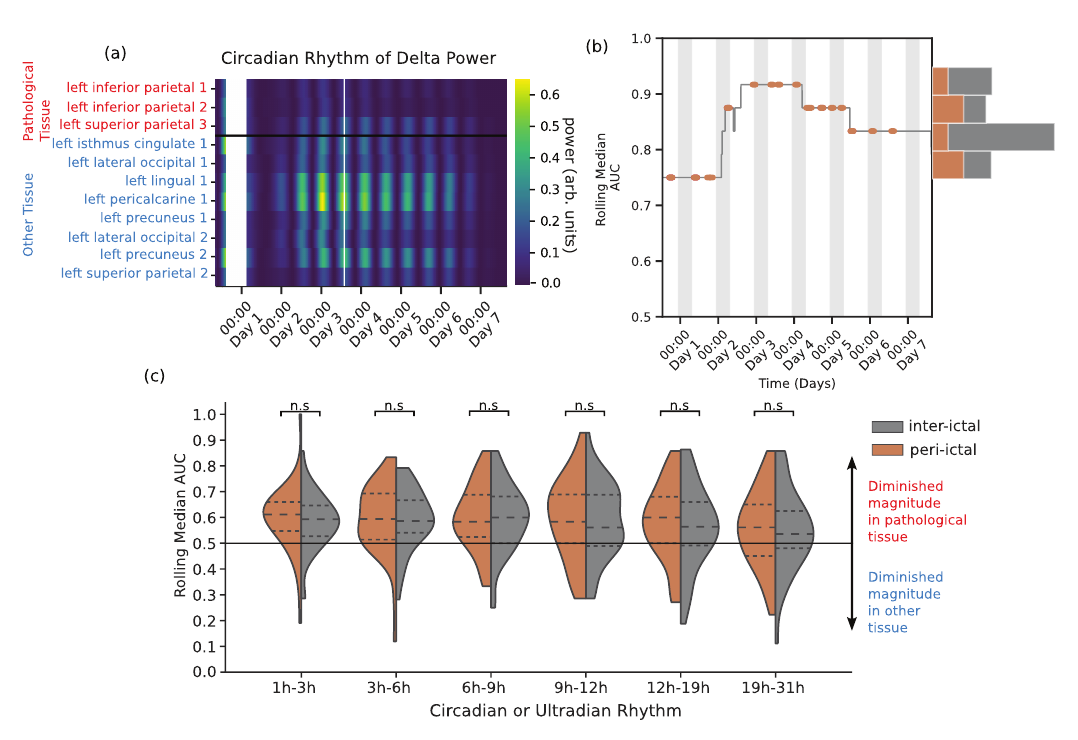}
    \caption{\textbf{Rhythms remain diminished in pathological tissue across time and independent of seizure occurrence.} a-e) Example subject A: (a) The power of the circadian rhythm of delta across ROIs. Each cell represents the power of the circadian rhythm in the corresponding ROI and 30 second time segment. ROIs in red are deemed pathological. (b) The median AUC of the circadian rhythm captured using a rolling window of 36 hours. The peri-ictal period (1 hour before seizure start to 1 hour after its end) is shown in orange. The histogram shows the distribution of AUC values calculated during peri-ictal and inter-ictal periods. (c) Compares rolling AUC values calculated during inter-ictal and peri-ictal periods for each circadian and ultradian rhythm of delta power. A two-sided paired Student's T test, with no correction for multiple comparisons indicates no difference between periods for each rhythm. Dashed lines indicate median and quartiles of the distribution.}
    \label{fig:fig4}
\end{figure}
%%%%%%%%%% %%%%%%%%%% SUPPLEMENTARY %%%%%%%%%% %%%%%%%%%% 

\renewcommand{\thefigure}{S\arabic{figure}}
\setcounter{figure}{0}
\counterwithin{figure}{section}
\counterwithin{table}{section}
\renewcommand\thesection{S\arabic{section}}
\setcounter{section}{0}

\newpage 

\section*{Supplementary}

%\url{https://newcastle-my.sharepoint.com/:f:/r/personal/b6071244_newcastle_ac_uk/Documents/cycles_disrupt_proj/AUC_by_pathology?csf=1&web=1&e=50aLej} shows Mariella's results by post-op pathology.

\section{Imputation of missing data\label{suppl:imputation}}

To facilitate further analysis, we employed imputation techniques to fill any gaps in the relative log band power of each ROI before proceeding with rhythm extraction at various timescales. However, we did not utilise imputed data for the final analysis. We identified missing blocks within each frequency band and imputed them accordingly. If a missing block had a size equal to one, we replaced it with the mean of the value before and after the block. For missing blocks larger than one, we identified the surrounding segments of equal length before and after the block. When the preceding segment was smaller than the missing block or the missing blocks were at the start of the recording, we used only the following segment for imputation. We interpolated the data of missing blocks using the mean of the adjacent segments and added Gaussian noise with a mean of zero and standard deviation of 60\% of the standard deviation of the surrounding segments. Any missing data present in the adjacent segments was disregarded. The final values used for analysis were the interpolated ones with added Gaussian noise.
\section{AUC distributions across rhythms and EEG bands\label{suppl:AUCtable}}

\begin{table}[!ht]
    \centering
        \caption{\textbf{Almost all AUC distributions are significantly greater than 0.5. Shows the T statistic and p value of a one-sided Wilcoxon signed rank test applied to test whether each distribution is greater than 0.5 for each EEG band/biological rhythm pair. No correction for multiple comparisons has been performed. }}
    \label{tab:figST1}
    \begin{tabular}{|l|l|l|l|l|l|}
    \hline
       ~ & \multicolumn{5}{|l|}{Wilcoxon rank sum test [\emph{T}, p]}  \\ \hline
        Rhythm & Delta  & Theta & Alpha & Beta & Gamma \\ \hline
1h-3h & [631, $<$0.001] & [538, $<$0.001] & [681, $<$0.001] & [662, $<$0.001] & [590, $<$0.001]  \\
3h-6h & [632, $<$0.001] & [486, 0.008] & [648, $<$0.001] & [649, $<$0.001] & [551, 0.004]  \\
6h-9h & [630, $<$0.001] & [568, 0.001] & [590, $<$0.001] & [630, $<$0.001] & [581, $<$0.001]  \\
9h-12h & [628, $<$0.001] & [493, 0.016] & [510, 0.009] & [550, $<$0.001] & [551, $<$0.001] \\
12h-19h & [576, 0.001] & [460, 0.051] & [490, 0.007] & [561, 0.001] & [511, 0.021]\\
19h-31h & [498, 0.005] & [479, 0.027] & [492, 0.006] & [594, $<$0.001] & [596, $<$0.001]\\ \hline
    \end{tabular}
\end{table}
\begin{table}[!ht]
    \centering
        \caption{\textbf{Almost all AUC distributions are significantly greater than 0.5. Shows the median AUC and the percent of AUCs above 0.5 for each EEG band/biological rhythm pair.}}
    \label{tab:figST2}
    \begin{tabular}{|l|l|l|l|l|l|}
    \hline
    ~ & \multicolumn{5}{|l|}{[median AUC, \% $AUC>0.5$]}  \\ \hline
    Rhythm & Delta & Theta & Alpha & Beta & Gamma  \\ \hline
     1h-3h  & [0.71, 82\%] & [0.64, 74\%] & [0.72, 87\%] & [0.7, 84\%] & [0.71, 74\%] \\
     3h-6h & [0.68, 71\%] & [0.65, 63\%] & [0.68, 82\%] & [0.69, 79\%] & [0.57, 63\%] \\
     6h-9h & [0.69, 76\%] & [0.68, 66\%] & [0.67, 76\%] & [0.75, 71\%] & [0.64, 71\%]  \\
     9h-12h & [0.66, 76\%] & [0.64, 63\%] & [0.62, 61\%] & [0.67, 68\%] & [0.62, 74\%] \\
     12h-19h & [0.61, 74\%] & [0.53, 53\%] & [0.61, 55\%] & [0.64, 68\%] & [0.56, 61\%]  \\
    19h-31h & [0.6, 71\%] & [0.55, 58\%] & [0.59, 68\%] & [0.67, 71\%] & [0.61, 76\%] \\\hline
    \end{tabular}
\end{table}

\begin{table}[!ht]
    \centering
    \caption{\textbf{Pathology implies diminished rhythms when controlling for brain region.} A mixed effects regression showing the effect of pathology on rhythm power. Here we show partially standardised coefficient for pathology (SOZ) for each mixed effects model predicting the cycle power from the brain region of the ROI, and whether the ROI is pathological (in the SOZ). The model is described in section \ref{sec:model} of the main text. Each model is built using data from 38 patients.}
\begin{tabular}{|l|l|l|l|l|l|l|l|}

\hline
EEG Band & Rhythm & SOZ Beta & SE & z & pvalue & lower95CI & higher95CI \\
\hline
Delta & 1h-3h       & -0.301 & 0.049 & -6.127 & $<$0.001 & -0.397 & -0.205 \\
Delta & 3h-6h       & -0.257 & 0.053 & -4.864 & $<$0.001 & -0.361 & -0.154 \\
Delta & 6h-9h       & -0.17 & 0.045 & -3.821 & 0.001 & -0.258 & -0.083 \\
Delta & 9h-12h      & -0.221 & 0.051 & -4.367 & $<$0.001 & -0.32 & -0.122 \\
Delta & 12h-19h     & -0.193 & 0.053 & -3.636 & 0.001 & -0.297 & -0.089 \\
Delta & 19h-1.3d    & -0.185 & 0.057 & -3.269 & 0.005 & -0.296 & -0.074 \\ \hline
Theta & 1h-3h       & -0.274 & 0.063 & -4.369 & $<$0.001 & -0.396 & -0.151 \\
Theta & 3h-6h       & -0.181 & 0.048 & -3.76 & 0.001 & -0.276 & -0.087 \\
Theta & 6h-9h       & -0.173 & 0.049 & -3.517 & 0.002 & -0.27 & -0.077 \\
Theta & 9h-12h      & -0.171 & 0.056 & -3.037 & 0.01 & -0.281 & -0.06 \\
Theta & 12h-19h     & -0.173 & 0.052 & -3.296 & 0.005 & -0.276 & -0.07 \\
Theta & 19h-1.3d    & -0.129 & 0.055 & -2.329 & 0.067 & -0.238 & -0.02 \\ \hline
Alpha & 1h-3h       & -0.262 & 0.072 & -3.662 & 0.001 & -0.402 & -0.122 \\
Alpha & 3h-6h       & -0.257 & 0.065 & -3.959 & $<$0.001 & -0.384 & -0.13 \\
Alpha & 6h-9h       & -0.196 & 0.058 & -3.352 & 0.004 & -0.311 & -0.081 \\
Alpha & 9h-12h      & -0.188 & 0.073 & -2.57 & 0.037 & -0.331 & -0.045 \\
Alpha & 12h-19h     & -0.191 & 0.068 & -2.836 & 0.018 & -0.324 & -0.059 \\
Alpha & 19h-1.3d    & -0.102 & 0.07 & -1.462 & 0.344 & -0.238 & 0.035 \\ \hline
Beta & 1h-3h       & -0.321 & 0.062 & -5.149 & $<$0.001 & -0.444 & -0.199 \\
Beta & 3h-6h       & -0.241 & 0.056 & -4.317 & $<$0.001 & -0.35 & -0.132 \\
Beta & 6h-9h       & -0.153 & 0.051 & -3.026 & 0.011 & -0.253 & -0.054 \\
Beta & 9h-12h      & -0.223 & 0.053 & -4.185 & $<$0.001 & -0.327 & -0.118 \\
Beta & 12h-19h     & -0.194 & 0.051 & -3.838 & 0.001 & -0.293 & -0.095 \\
Beta & 19h-1.3d    & -0.236 & 0.058 & -4.052 & $<$0.001 & -0.35 & -0.122 \\ \hline
Gamma & 1h-3h       & -0.181 & 0.037 & -4.847 & $<$0.001 & -0.254 & -0.108 \\
Gamma & 3h-6h       & -0.156 & 0.044 & -3.582 & 0.002 & -0.241 & -0.071 \\
Gamma & 6h-9h       & -0.186 & 0.04 & -4.599 & $<$0.001 & -0.266 & -0.107 \\
Gamma & 9h-12h      & -0.198 & 0.042 & -4.68 & $<$0.001 & -0.281 & -0.115 \\
Gamma & 12h-19h     & -0.205 & 0.046 & -4.487 & $<$0.0010.0 & -0.294 & -0.115 \\
Gamma & 19h-1.3d    & -0.216 & 0.051 & -4.201 & $<$0.001 & -0.317 & -0.115 \\
\hline
\end{tabular}

\end{table}
\section{Diminished rhythms and seizure load\label{suppl:szload}}

\begin{figure}
    \centering
    \includegraphics[width=0.7\linewidth]{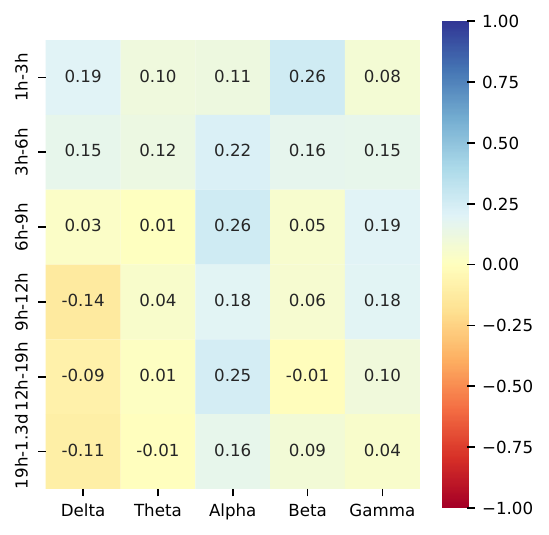}
    \caption{\textbf{There are no strong correlations between seizure load and diminished chronobiological rhythms.} The seizure load was calculated as the number of seizures per hour. A Pearson correlation was calculated between this and the AUC for each ultradian and circadian rhythm and EEG band. All coefficients were below 0.3 (with most below 0.1), indicating that in most cases there was no correlation and at most a weak correlation. Above figure shows the coefficients (r) for each combination. For each correlation n=38.   }
    \label{fig:figS1}
\end{figure}

In the main paper we show than the diminished rhythms is consistent in time, without dependence on seizure timing. To provide further evidence that seizure load is not driving the result, we here show (figure \ref{fig:figS1} the correlation between seizure load and AUC. We found either no correlation or weak correlation between seizure load and AUC.

% \section{AUC distributions in faster ultradians and infradians\label{suppl:extraUltr}}

% Here we show distribution of AUC values for ultradian rhythms less than 1 hour and infradian rhythms. 
% \begin{figure}
%     \centering
%     \includegraphics[width=1\linewidth]{SOZ_allcyclesAUCdists.pdf}
%     \caption{AUC distributions for each EEG band and a greater range of biological rhythms than presented in the paper.}
%     \label{fig:S2}
% \end{figure}

\section{Signal magnitude in the SOZ\label{suppl:rawmag}}

We wished to rule out the possibility that differences in the power of the signal between the SOZ and the rest of the tissue may be driving the diminished rhythms we see. To investigate this we have summed the raw band power values in all frequency bands and across time to obtain the overall signal power in each ROI (equation \ref{eq:totalpw}). Where $T$ is the number of samples, $\delta BP$ is a (T x nROI) matrix containing the power of the signal in the delta band, and TotalPowerROI is a vector containing the total power of the signal in each ROI. We then calculated an AUC with the $TotalPowerROI$ in ROIs within the SOZ against the $TotalPowerROI$ in ROIs not in the SOZ to get an $AUC_{TP}$. 
\begin{equation}
    TotalPowerROI = log \left (\sum_{t=0}^{t=T} \delta BP_{t} + \theta BP_{t} + \alpha BP_{t} + \beta BP_{t} + \gamma BP_{t} \right)
    \label{eq:totalpw}
\end{equation}
\begin{equation}
    AUC_{TP} = AUC \left ( TotalPowerROI[SOZ], TotalPowerROI[notSOZ] \right)
    \label{eq:totalAUC}
\end{equation}

\begin{figure}
     \centering
     \includegraphics[width=0.6\linewidth]{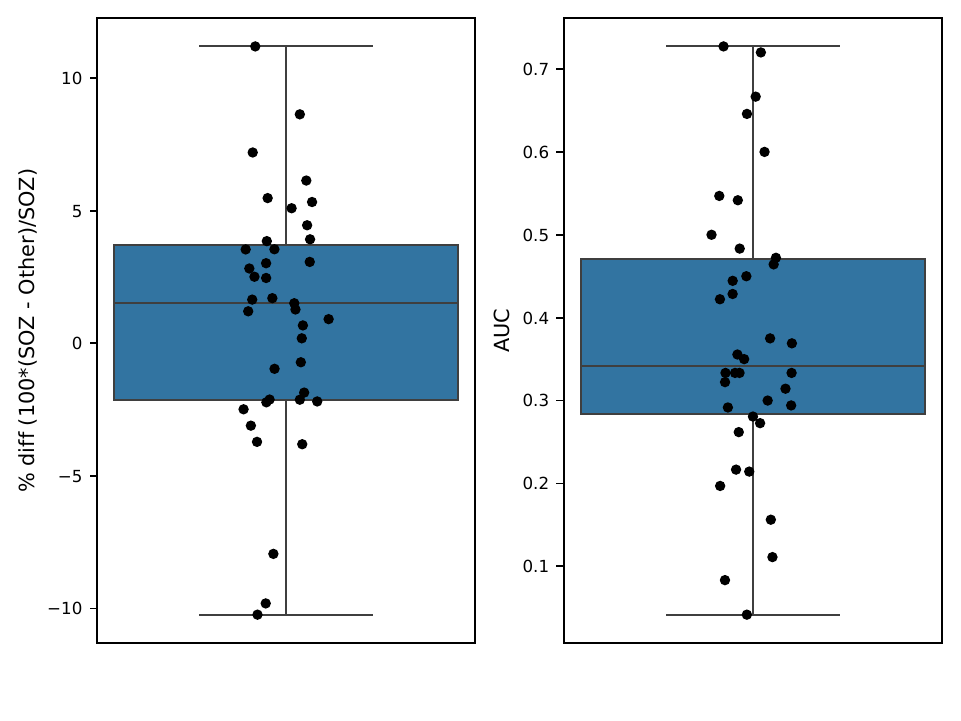}
     \caption{(A) The difference between the raw log power in the SOZ and non-SOZ. (B) Using the raw log power to calculate an AUC in the same manner as we calculated AUCs for the power of the chronobiological rhythms. }
     \label{fig:S3}
\end{figure}

We first tested whether the log of the overall signal power differed in pathological tissue (SOZ) compared to other tissue (Figure \ref{fig:S3} a). In terms of percentage difference between pathological and other tissue, there was a wide spread across the cohort from -20\% to +20\%, and the median as approx +3\% indicating little to no difference in overall signal power between tissue types on average. If at all, the pathological tissue may show elevated signal power across the cohort, possibly due to interictal spike or similar. We also tested if overall signal power distinguishes pathological tissue (SOZ) from other tissue using the $AUC_{TP}$ (Figure \ref{fig:S3} b). Over the cohort, again, there was a slight shift towards pathological tissue displaying higher overall signal power than other tissue, again, possibly due to interictal phenomena.

\begin{figure}
     \centering
     \includegraphics[width=1\linewidth]{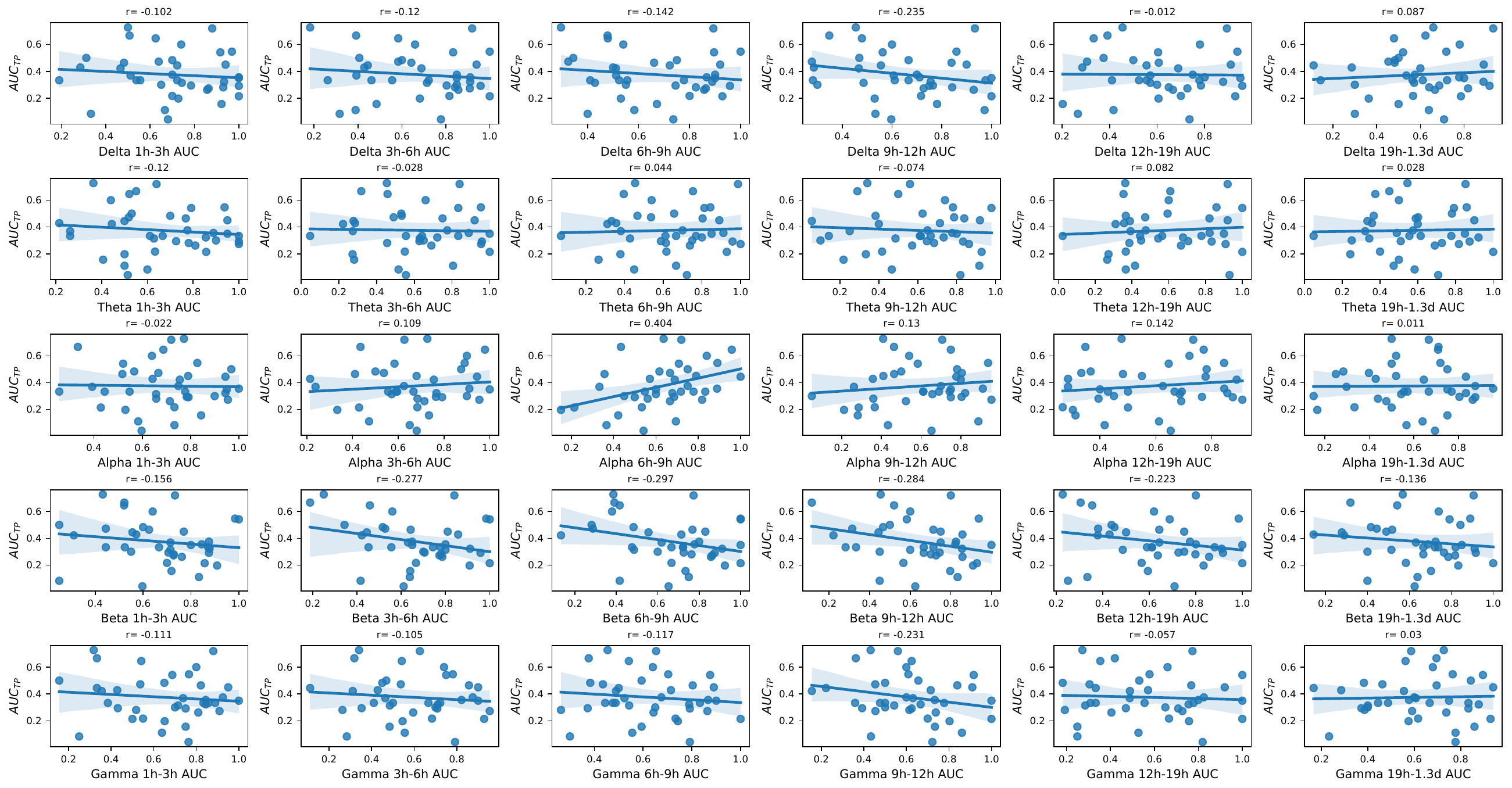}
     \caption{The correlation between the raw power and the AUC calculated on the chronobiological rhythm for each EEG band. For delta, theta, and alpha, we see no correlation or minimal correlation (r$<$0.3). For beta and gamma moderate negative correlations are present. }
     \label{fig:S4}
 \end{figure}

We then directly tested if $AUC_{TP}$ correlated with the AUC values we obtained in the main results (using the relative band powers in each band). Our results show that this is not the case, most correlations are very weak ($<0.2$), and all correlations are $<0.4$ (Figure \ref{fig:S4}). We conclude that the chronobiological rhythms seen in relative band power are not directly determined by overall signal magnitude, i.e. the reported effects are not simply due to a weaker EEG signal in pathological tissue.


\section{Cycles are also diminished in resected \textit{vs.} spared tissue\label{suppl:resvsspared}}

In the main analyses of this paper we have used the seizure onset zone (SOZ) to indicate whether pathology is present in a region. An alternative marker is whether the tissue was resected during surgery. To assess whether chronobiological rhythms remained diminished when using a resection as a marker for pathology, we re-calculated AUC values for each rhythm and EEG band using this instead of SOZ, and present them above. For delta, theta, alpha, and beta, we find median AUC values above 0.5 for most rhythms - similar to the results using SOZ and indicating diminished rhythms in resected tissue. For gamma, we median AUC values around 0.5 - indicating no difference in the strength of the chronobiological rhythms between resected and spared tissue.
\begin{figure}
    \centering
    \includegraphics[width=0.9\linewidth]{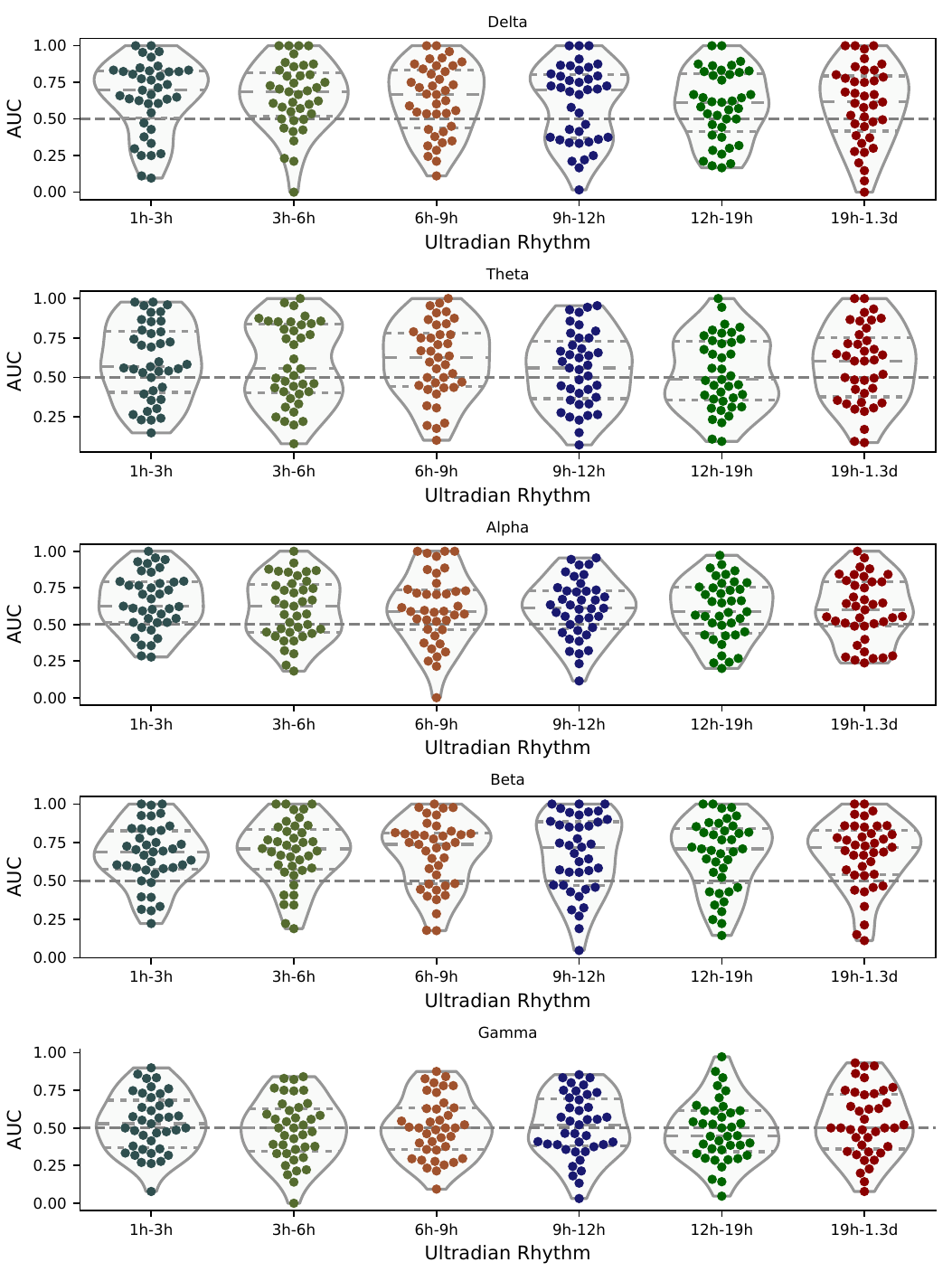}
    \caption{The AUC values calculated on resected regions against spared regions for each patient in each chronobiological rhythm and each EEG band.}
    \label{fig:S5}
\end{figure}

\section{Cycles are diminished in multiple types of pathology\label{suppl:pathtype}}
\begin{figure}
    \centering
    \includegraphics[width=1\linewidth]{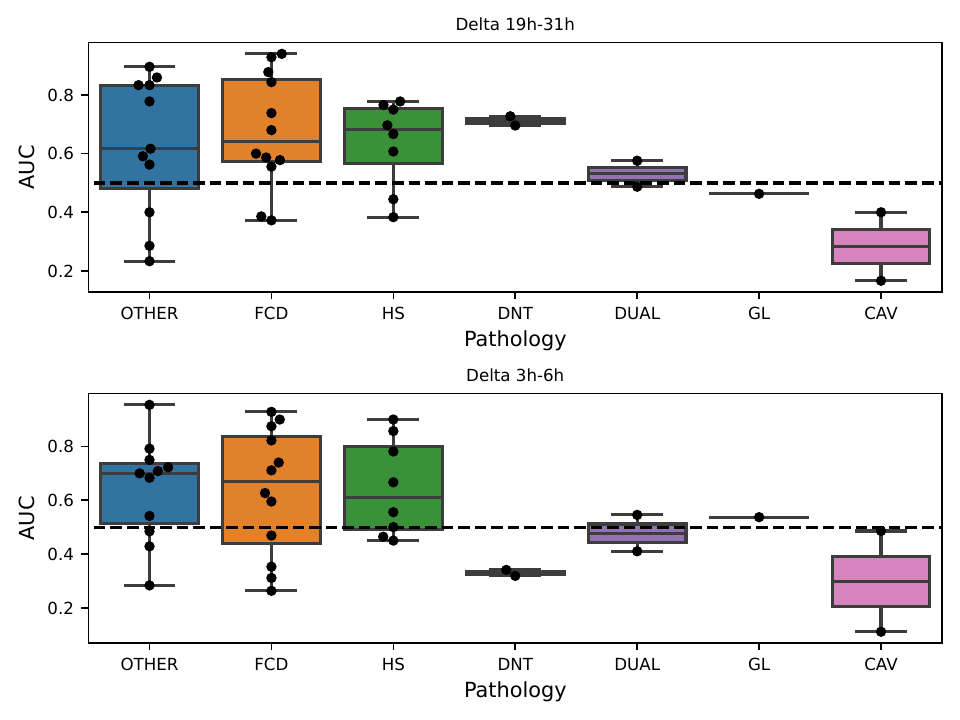}
    \caption{Top: AUC values of distinguishing pathological tissue \textit{vs.} other tissue in terms of the circadian rhythm magnitude in relative delta band power plotted by type of pathology. Bottom: Same as top, but for the 3-6h ultradian rhythm. Pathology abbreviations: hippocampal sclerosis (HS); focal cortical dysplasia (FCD); dysembryoplastic neuroepithelial tumor (DNT); gliosis (GL); cavernoma (CAV); multiple pathologies (DUAL), other pathologies (OTHER)}
    \label{fig:pathology}
\end{figure}

Figure \ref{fig:pathology} shows the AUC of distinguishing pathological tissue from other tissue in terms of magnitude in the circadian and one example ultradian rhythm in relative delta band power. Our sample was too small to draw definitive conclusion on any individual pathology, but the most common pathologies of hippocampal sclerosis (HS) and focal cortical dysplasia (FCD) both demonstrate the diminished rhythmicity. Also most subjects show this effect, without sufficient evidence for an influence of pathology type.

\section{Diminished rhythms across age, sex, epilepsy type, and anti-seizure medication}
\label{suppl:agesexmed}
\begin{figure}
    \centering
    \includegraphics[width=1\linewidth]{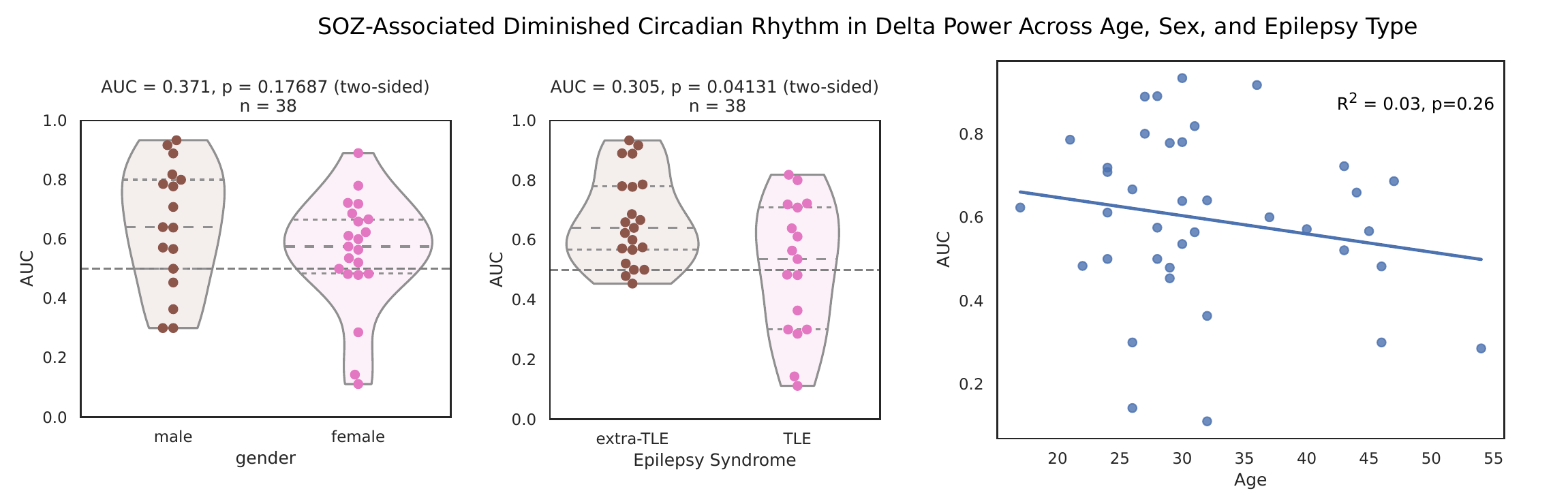}
    \caption{AUCs indicating the diminished power of the circadian rhythm in delta power in the SOZ, comparing patients split by Sex, Epilepsy Type (TLE or eTLE), and age. We saw no significant difference with regards to sex, but the rhythm power was more diminished in eTLE compared to TLE (p=0.04). A negative slope with found in the regression with age, however this was weak ($R^2=0.03$).   }
    \label{fig:agesexep}
\end{figure}

\begin{figure}
    \centering
    \includegraphics[width=1\linewidth]{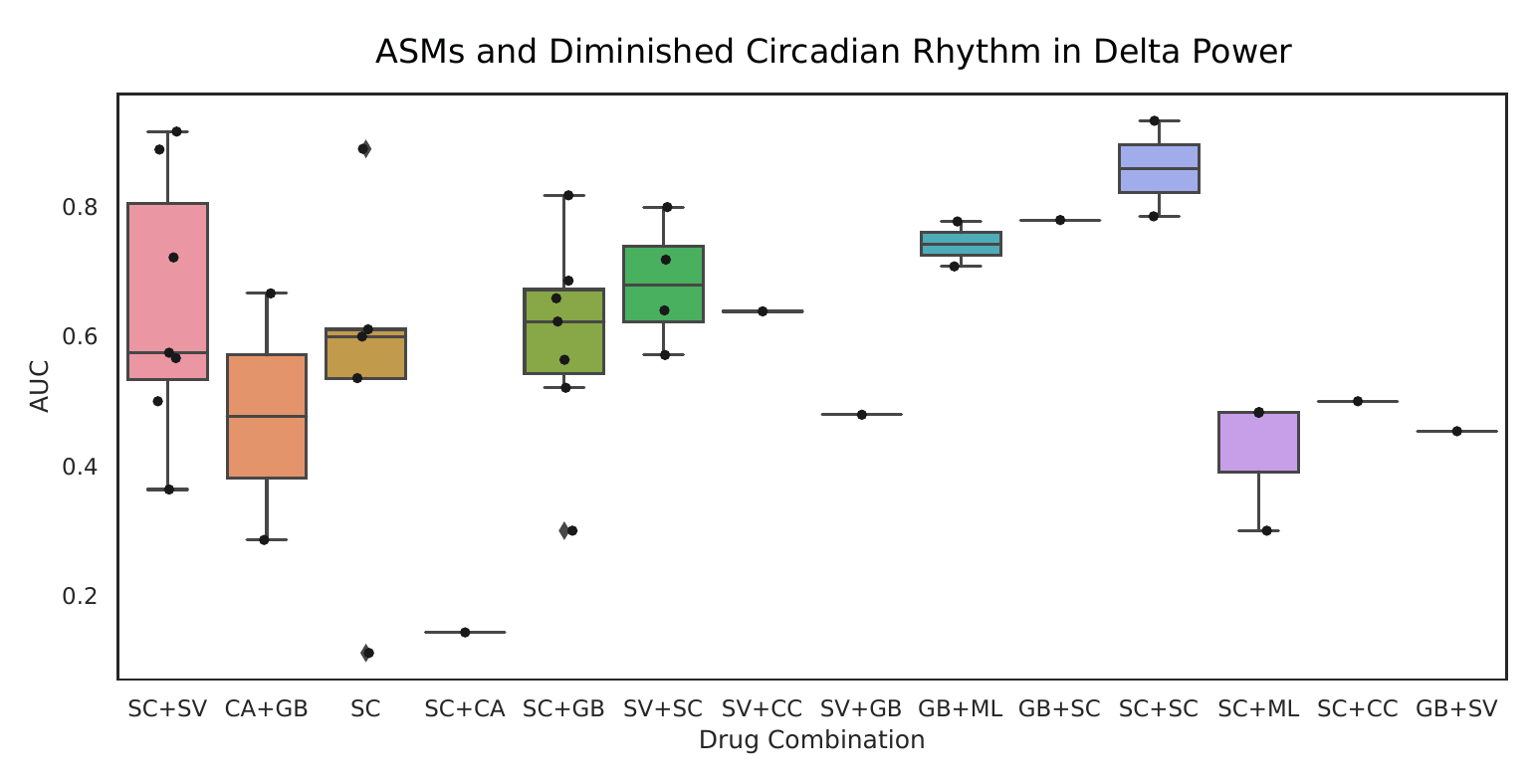}
    \caption{AUC indicating the diminished power of the circadian rhythm in delta power in the SOZ, comparing the anti-seizure medication (ASM) combinations the patients were using at the time. Due to the large variation in drugs used, and the multiple combinations of drugs, we could not draw any conclusions from this data. Drug type is indicated by its mechanism of action, with SC: sodium channel blocker (e.g. Carbamazapine, Lamotrigine), GB: GABA enhancer (e.g. Clonazepam, Diazepam), CC: calcium channels (e.g. Gabapeptin), GT: glutamate receptors (e.g. Perampanel), SV: SV2a receptor (e.g. Levitiracitam), CA: inhibition of carbonic anhydrase (e.g. Acetazolamide), ML: multiple targets (e.g. Sodium Valproate).  }
    \label{fig:drugs}
\end{figure}

Overall, we did not observe noteworthy modulatory/differential effects of age, sex, epilepsy type, or anti-seizure medication on the diminished power of the circadian rhythm in pathological tissue (Fig.~\ref{fig:agesexep} and Fig.~\ref{fig:drugs}). We did observe a moderate difference (with $AUC=0.3$ and $p=0.04$) between TLE and eTLE in terms of their AUC distinguishing signal power between pathological and other regions. However, upon further investigation by including a categorical variable into out mixed effect model, we did not see any substantial or significant effects on TLE vs eTLE. Therefore, the moderate effect seen here is most likely driven by the different regions implanted in TLE vs eTLE.

\section{Using a tighter definition of circadian rhythms}
\label{suppl:circ26}
\begin{figure}[ht!]
    \centering
    \includegraphics[width=1\linewidth]{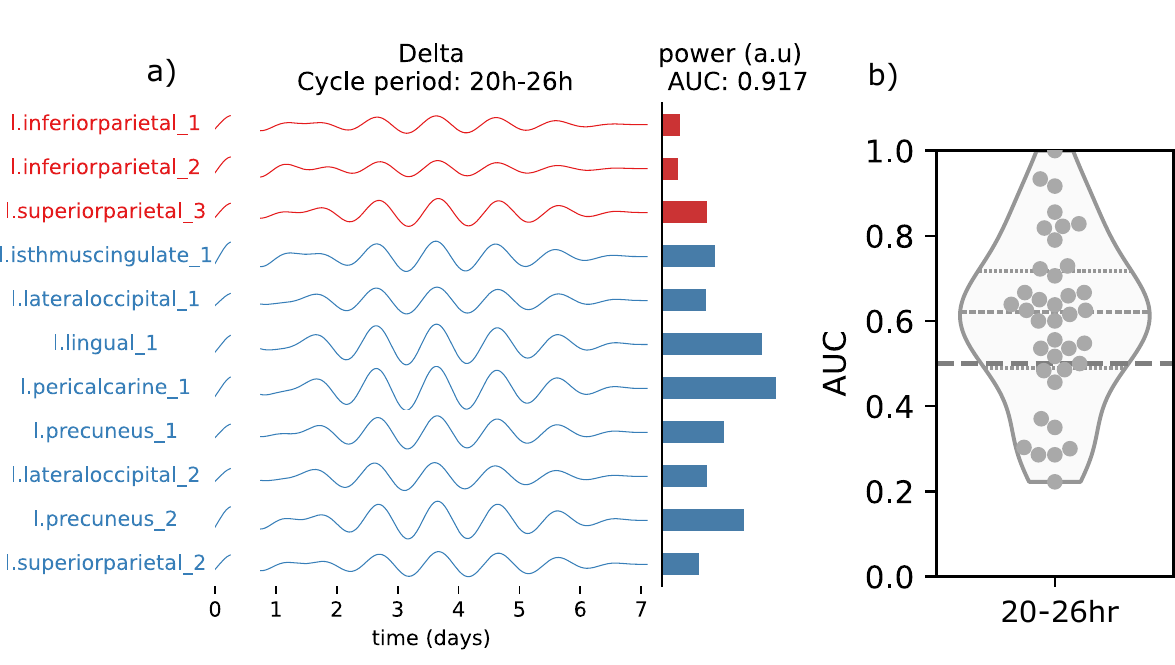}
    \caption{Circadian rhythms are diminished in the SOZ to a similar extent when using a narrower definition. a) Shows the power of delta oscillations in each region, filtered to isolate the circadian component (with a period of between 20 and 26 hours) for the example subject used in figure \ref{fig:fig1} in the main text. Red lines indicate regions that are pathological (within the SOZ). This subject has an AUC of 0.917 indicating that the magnitude of the rhythm is diminished in the pathological areas.  b) shows the value of AUCs for all patients, most are above 0.5 indicating that the rhythm is diminished in the SOZ for most of the cohort.}
    \label{fig:circ26}
\end{figure}

In the main text we use a broad definition of circadian rhythms as fluctuations with a period of between 19 and 31 hours, and parameterise our band pass filter accordingly. To investigate whether a narrower definition of the circadian fluctuations we recalculate them using a definition of 20-26 hours. Figure \ref{fig:circ26} replicates figure \ref{fig:fig1} (a and d) from the main text, using the narrower definition. We find that for our example subject we get a similar pattern of signal power across regions, and we get a similar overall AUC of 0.917. Figure \ref{fig:circ26} (b) shows that across the population we get a very similar distribution of AUCs, with 0.6 as the median.  

\section{Circadian patterns of seizure occurrence}
\label{suppl:phasepref}
\begin{figure}[ht!]
    \centering
    \includegraphics[width=1\linewidth]{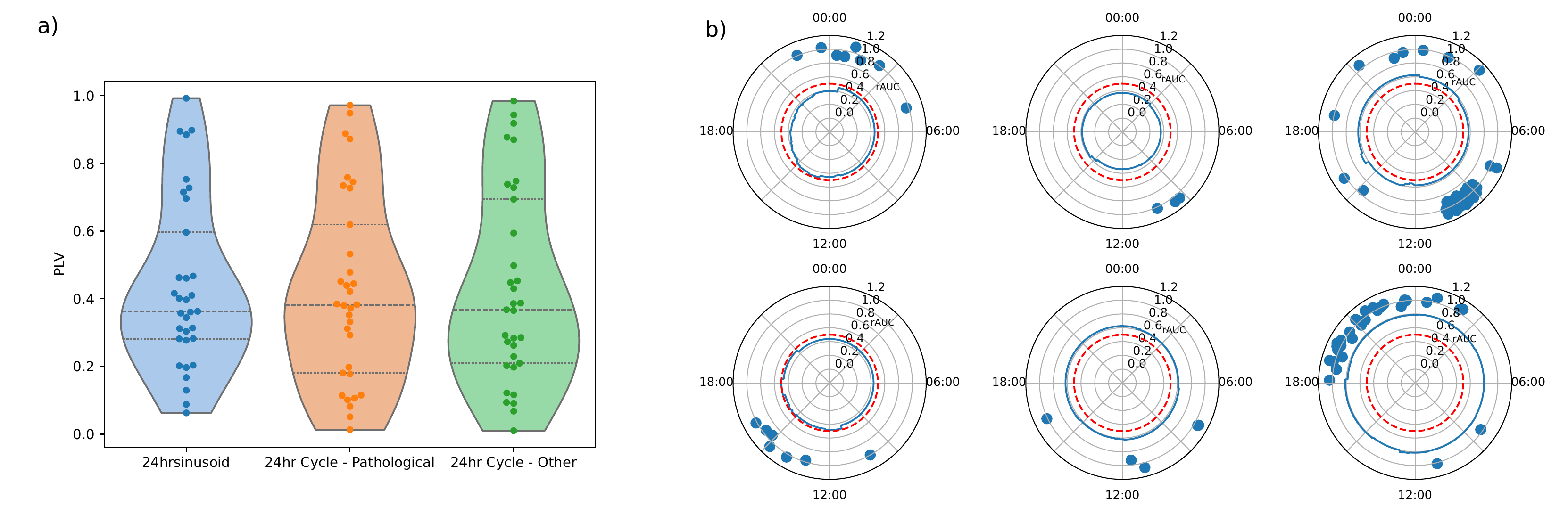}
    \caption{a) Shows the distribution of phase locking values across the cohort quantifying the phase locking of their seizures to the 24 daily clock cycle, the circadian cycle of their delta power in pathological brain regions and their circadian cycle in delta power in their other brain regions. b) Shows the timings of seizures in the 24 hour cycle for the 6 patients with the highest phase-locking value - blue circles. The rolling AUC (rAUC) is plotted in blue with the dashed red line indicating an AUC of 0.5. }
    \label{fig:phase_pref}
\end{figure}

While this study focussed on the circadian and ultradian rhythmicity in the power of the canonical frequency bands in the iEEG, it is well known that seizures themselves can occur at particular phases in these rhythms for some patients. To investigate further whether this played any role in the diminished rhythms we have found in the SOZ, we show here the phase locking value of seizures to the 24 hour daily cycle, the filtered circadian cycle in delta power in the pathological regions, and in the other regions. Phase locking value (PLV) here is calculated following previously outlined methods \citep{Schroeder2023}, where a PLV of 1 indicates all seizure occur at the same phase, while 0 indicates a uniform distribution of phases. We found that for the majority of patients the phase locking values did not imply strong phase preference for any of the three cycles. In a minority of patients phase preference did exist, for the 6 patients with the strongest preference in figure \ref{fig:phase_pref} (b) we show the mean rAUC across the 24 hour cycle and the phase of each seizure. In none of these cases do we see changes in the rAUC at the phase preference of the seizures, emphasising that rhythms are diminished independently of seizure occurrence.  

\section{AUCs in the post-ictal period}
\label{suppl:post_ictal}
\begin{figure}[h]
    \centering
    \includegraphics[width=0.75\linewidth]{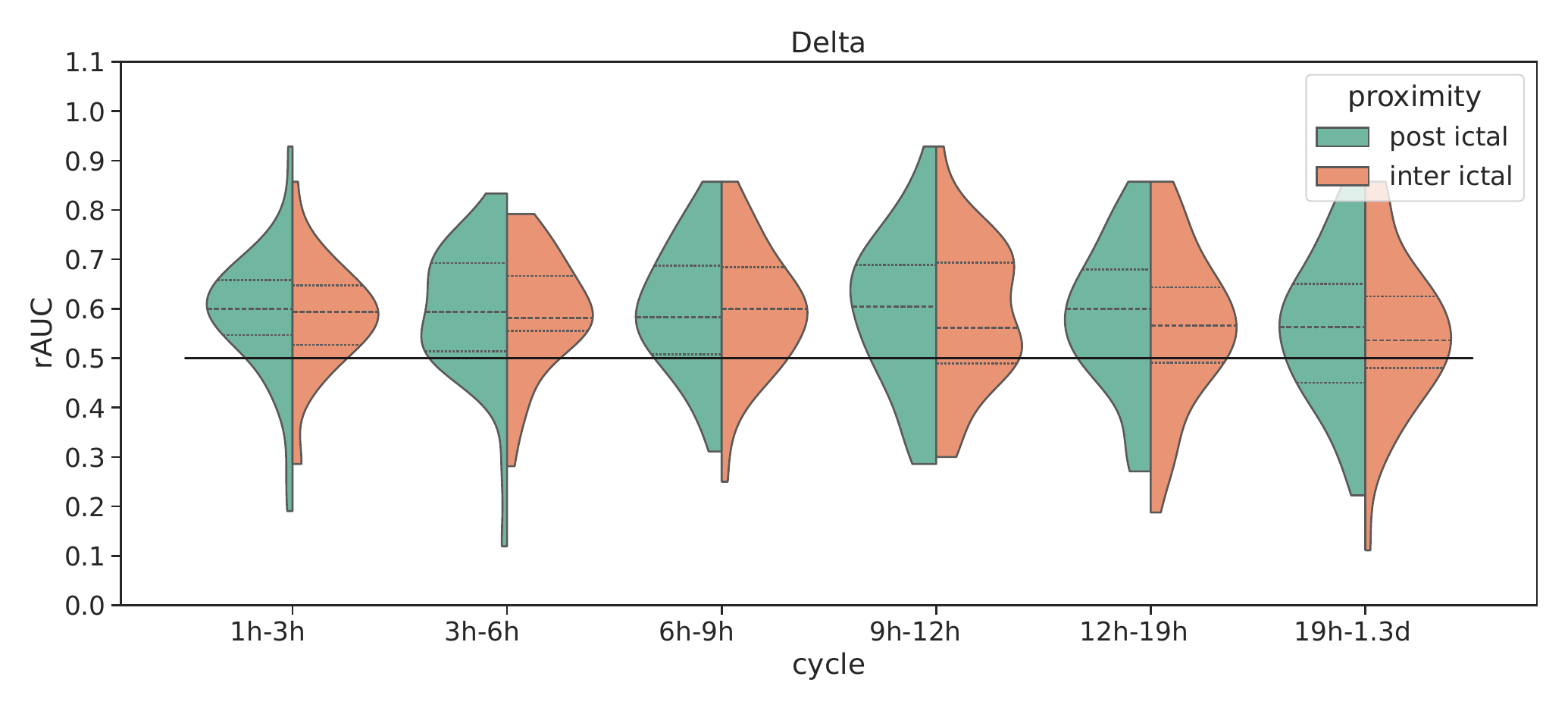}
    \caption{Shows the rAUCs calculated using only the power of the daily rhythm in delta power in the 30 minutes post-seizure.}
    \label{fig:postictal}
\end{figure}

While we show the AUC values in the peri-ictal period in figure \ref{fig:fig4}, the post-ictal period - 30 minutes post seizure - is also known to show aberations in the EEG. Here we calculated an rAUC value solely based on the power of the circadian rhythm during this 30 minute period, shown in figure \ref{fig:postictal}. We find that the AUC values here are similar to those presented in figure \ref{fig:fig4}. 

\section{Changes in daily rhythm power over time}
\label{suppl:over_time}
\begin{figure}
    \centering
    \includegraphics[width=\linewidth]{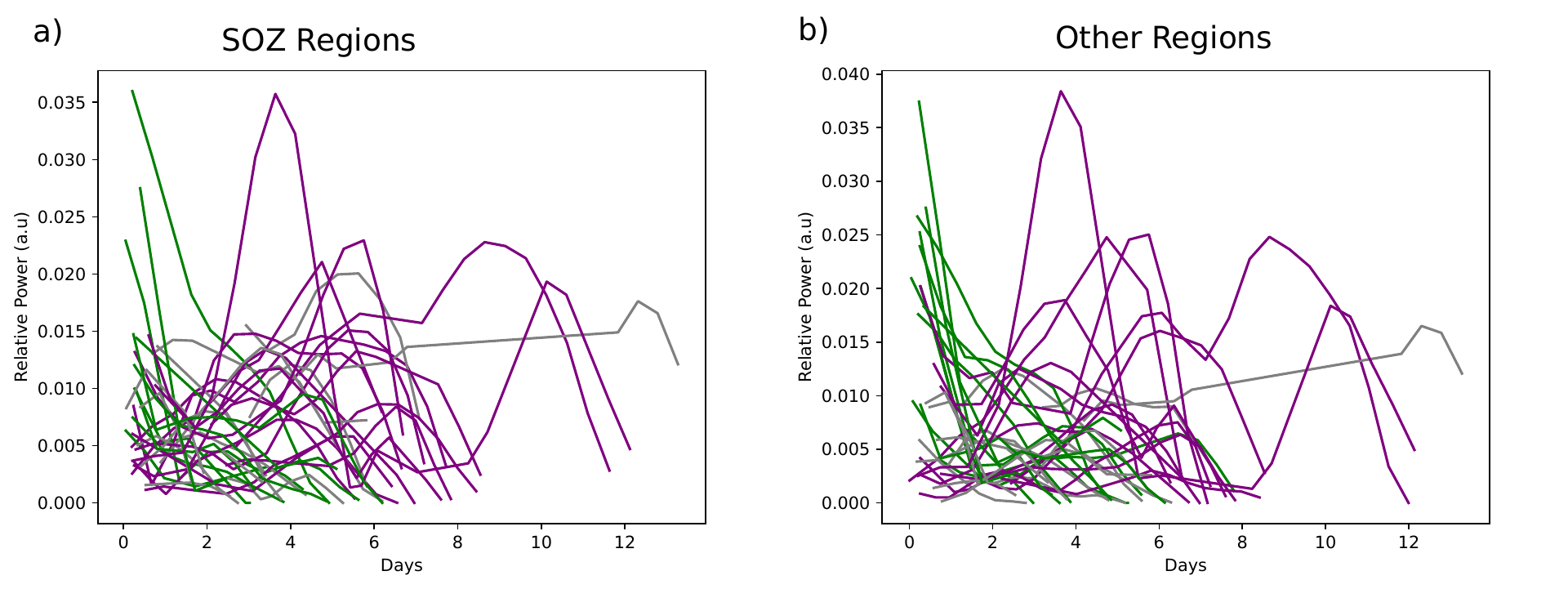}
    \caption{Shows the amplitude of the daily (circadian) rhythm in delta power over time for each of the patients, in the SOZ (a) and in other regions (b). Colors correspond to three categories. Green - clear decrease in amplitude (the maximum peak occurs in the first three peaks), purple - fits well to a Gaussian curve ($p<0.01$ in a one-sided Kolmogorov-Smirnov test), and the remainder of patients.}
    \label{fig:ovtime}
\end{figure}

The amplitude of the biological rhythms does not remain constant across the recording period for most patients. To investigate this we captured the amplitude of the daily rhythm in delta power over time for each patient by finding the peaks of the absolute value of the band pass filtered signal. We did this for signals within the SOZ and for those in other brain regions. This gave us the lines shown in figure \ref{fig:ovtime}. We then assigned these to three categories - those that show a clear amplitude loss over time (SOZ n=12, other n=10); those that fit well to a Gaussian curve, with a peak in the middle of the recording and lower amplitudes at the edges (SOZ n=13, other n=15); and those that do not meet either of these criteria (SOZ n=14, other n=14). 
\end{document}